\documentclass[a4paper,11pt]{article}

\usepackage{jcappub} 

\usepackage[T1]{fontenc} 
\usepackage{natbib}
\usepackage{amssymb}
\usepackage{amsmath}
\usepackage{amsfonts}
\usepackage{graphicx}
\usepackage[hang,tight,raggedright]{subfigure}
\usepackage{multirow}
\usepackage{subfigure}
\usepackage{float}
\usepackage{graphics}
\usepackage{slashed}
\usepackage{color}
\usepackage{bm}
\usepackage{braket}
\usepackage{psfrag}
\usepackage{booktabs}
\usepackage{latexsym}
\usepackage{pythonhighlight}

\usepackage{setspace}
\usepackage{amsmath}
\usepackage{amsthm}
\usepackage{mathrsfs}
\usepackage{epsf}
\usepackage{epsfig}
\usepackage{bbm}
\usepackage{bm}
\usepackage{amsfonts}
\usepackage{amssymb}
\usepackage{latexsym}
\usepackage[english]{babel}
\usepackage{multirow}
\usepackage{float}
\usepackage{url}
\usepackage{slashed}
\usepackage{xcolor} 
\usepackage[utf8]{inputenc}
\usepackage{stmaryrd} 
\usepackage{enumitem}
\usepackage{siunitx}
\usepackage{verbatim}
\usepackage{bbm}

\usepackage{caption3}
\usepackage{appendix}

\makeatletter

\newcommand{\Rmnum}[1]{\expandafter\@slowromancap\romannumeral #1@}
\makeatother

\newcommand{\be}{\begin{equation}}
	\newcommand{\ee}{\end{equation}}
\newcommand{\ba}{\begin{array}}
	\newcommand{\ea}{\end{array}}
\newcommand{\bea}{\begin{eqnarray}}
	\newcommand{\eea}{\end{eqnarray}}

\usepackage{theorem}
{
	\theoremheaderfont{\bfseries}
	\theorembodyfont{\normalfont}
}

\title{Freeze-in Gravitational Waves and Dark Matter in Warm Inflation}
\author[a]{Quan Chen,}
\author[a]{Siyu Jiang,}
\author[a]{Dayun Qiu,}
\author[a]{Peilin Chen,}
\author[a,1]{and Fa Peng Huang\note{Corresponding author.}}
\emailAdd{huangfp8@sysu.edu.cn}
	
\affiliation[a]{MOE Key Laboratory of TianQin Mission, TianQin Research Center for Gravitational Physics \& School of Physics and Astronomy, Frontiers Science Center for TianQin, Gravitational Wave Research Center of CNSA, Sun Yat-sen University (Zhuhai Campus), Zhuhai 519082, China}

\abstract{

Recent study~\cite{Berghaus:2025dqi} has suggested that warm inflation may be realized with a minimal extension of the Standard Model by a single scalar inflaton field with an axion-like coupling to gluons. Motivated by this framework, we investigate the  gravitational wave spectrum and graviton-portal dark matter production through the freeze-in process generated during warm inflation scenarios. 
We perform a comparative analysis for different dissipation terms, focusing on their distinct gravitational wave signatures in the high-frequency regime. Our findings reveal qualitative and quantitative differences in the spectral behavior, offering a preliminary pathway for exploring inflationary and dark matter models through high-frequency gravitational wave signals.}


\begin{document}
\maketitle
\flushbottom

\section{Introduction}
Cosmic inflation~\cite{STAROBINSKY198099,Guth1981,LINDE1} can naturally explain the cosmic microwave background radiation and the large-scale structure observations, since the density perturbation of the inflaton field can provide the primordial seeds of our Universe. Warm inflation (WI)~\cite{WI1,WI2,WI3,WI4,WI5} is one type of inflationary scenario that fundamentally alters the conventional picture of the early Universe's expansion by including dissipative effects and concurrent radiation production during inflation. Unlike the standard cold inflation (CI) model, where the Universe supercools and reheating occurs from the late stage of inflation~\cite{huang2025particles}, WI maintains a thermal bath throughout the inflation process by dissipating the inflaton field's energy into radiation. This continuous particle production modifies the dynamics of the inflaton field and the generation of primordial density perturbations, offering possible solutions to challenges faced by CI such as the “eta problem” and allowing for sub-Planckian field excursions.
The conceptual foundation of WI integrates quantum field theory techniques to model dissipative processes, showing that interactions between the inflaton field and other fields produce radiation simultaneously with inflationary expansion. 

Although the concept of WI presents a compelling alternative to conventional cold inflationary scenarios, constructing a theoretically consistent WI model has proven challenging. Early formulations of WI relied on scalar couplings between the inflaton field and the particles comprising the thermal bath. However, subsequent analyses revealed that scalar interactions predominantly induce thermal corrections to the inflaton potential, rather than generating significant dissipative friction. These thermal effects tend to hinder inflationary dynamics rather than facilitate them, complicating efforts to build viable WI models based on scalar field interactions~\cite{Yokoyama:1998ju}.


Recently, K.V. Berghaus, M. Drewes, and S. Zell in ref.~\cite{Berghaus:2025dqi}, demonstrate, for the first time, that WI might be consistently realized within a minimal extension of the Standard Model, incorporating a scalar inflaton field with an axion-like coupling to gluons and a monomial potential. The CMB predictions and viability of this model have been further investigated in~\cite{SMWI2}. This finding might provide a possible framework for dissipative dynamics during inflation and establishes a concrete connection between early-Universe cosmology and well-motivated particle physics interactions. Motivated by this idea in ref.~\cite{Berghaus:2025dqi}, we investigate two cosmic relics produced in WI scenario: (i) the gravitational waves (GWs) originating from the scattering of thermal plasma constituents, and (ii) the graviton-portal dark matter (DM). Among the many potential sources of primordial GWs~\cite{Preheat1997,PreheatGW2,PreheatGW2007,PreheatGW2008,ReheatGW2019,ReheatGW2,Barman:2023ymn,Datta:2024tne,Datta:2025wfh,Xu:2025wjq,defect1,defect2,defect3,phase1,phase2,phase3,phase4,phase5,phase6,turbulence1,turbulence2,turbulence3,turbulence4},  GWs generated by scattering processes among thermal plasma particles~\cite{Ghiglieri:2015nfa,Ghiglieri:2020mhm,Ringwald:2020ist,Ghiglieri:2022rfp,Drewes:2023oxg,2graviton,Montefalcone:2025gxx} are guaranteed to be present since it relies solely on the presence of a thermal bath. In this work, we refer to this GWs as ``freeze-in graviton'' or ``freeze-in GWs''~\cite{Ghiglieri:2022rfp} since the graviton and the graviton-portal DM production mechanism closely resembles the freeze-in DM scenarios~\cite{UVDM,Freese:2024ogj,WIFIDM,Ghiglieri:2022rfp}. The production of freeze-in particles is primarily determined by the thermal history of the early Universe; we therefore anticipate that it can effectively probe WI models, whose thermal history differs significantly from that of CI. In WI, the Universe naturally transitions from the inflationary epoch to the radiation-dominated era without requiring a separate reheating phase, allowing for the continuous production of gravitons and DM. In this work, we mainly investigate the freeze-in GW spectrum in the high-frequency regime and the associated DM production mediated by gravitons, within the framework of WI. Our analysis encompasses both a simplified version of the recently proposed WI scenario based solely on Standard Model gauge interactions~\cite{Berghaus:2025dqi}, as well as the typical WI model considered in ref.~\cite{Montefalcone:2025gxx}.

This paper is organized as follows. In section~\ref{WI}, we briefly review the mechanism of WI. In section~\ref{FIPG}, we analyze the freeze-in mechanism of gravitons in WI, with a review of the analogous DM WI freeze-in provided in appendix~\ref{appA}. In section~\ref{SPECTRA}, we calculate the GW spectrum for different WI models. In section~\ref{FIPPGDM}, we also estimate DM production solely by gravitational interaction from Standard Model plasma, and study the correlation between the DM mass and the peak amplitude of GW spectra. Finally, a concise conclusion is given in section~\ref{Conculsions}.

\section{Warm inflation}\label{WI}
In this section, we briefly review the background dynamics for general WI models.
The evolution of the background is governed by the following equations:
\begin{equation}
    \ddot{\phi}+(3H+\Upsilon)\dot{\phi}+V'(\phi)=0
\end{equation}
\begin{equation}
    \dot{\rho}_r+4H\rho_r=\Upsilon\dot{\phi}^2
\end{equation}
\begin{equation}
   H^2= \frac{8\pi}{3M_{\rm pl}^2}\left(\frac{\dot{\phi}^2}{2}+V(\phi)+\rho_r\right)  
\end{equation}
where $\phi$ is the inflaton field, $H$ the Hubble parameter, $\rho_r$ the radiation energy density, and $M_{\rm pl}\simeq1.221\times10^{19}~\mathrm{GeV} $ the Planck mass, respectively. The dot notation refers to the derivative with respect to cosmic time $t$, while the prime refers to the derivative with respect to the inflaton field $\phi$. Here, $\Upsilon$ is the dissipation coefficient which decides the energy transfer between the inflaton and the thermal bath. The distinction between the WI and CI is that in the former model, the dissipation term causes the radiation energy not to vanish. In this work, we assume the inflation potential 
\begin{equation}
    V(\phi) = \frac{\lambda}{4}\phi^4
\end{equation}
The radiation energy density with temperature $T$ is given by
\begin{equation}
    \rho_r=\frac{\pi^2}{30}g_*T^4
\end{equation}
where $g_*$ is the degrees of freedom of the thermal bath and we set $g_*=106.75$.

For convenience, one defines the dimensionless dissipation ratio $Q\equiv\frac{\Upsilon}{3H}$, which gives a measure of the strength of the dissipation effects in comparison to the Hubble expansion. Then, we can classify WI based on the parameter $Q$ as follows: $Q>1$ marks the strong dissipative regime while $Q<1$ marks the weak dissipative regime.

The slow-roll parameters in WI are modified to be
\begin{equation}
    \epsilon_V\equiv\frac{M_{\rm pl}^2}{16\pi(1+Q)}\frac{V'}{V}\quad,\quad\eta_V\equiv\frac{M_{\rm pl}^2}{8\pi(1+Q)}\frac{V''}{V}
\end{equation}
And during inflation, we have
\begin{equation}
    \epsilon_V\simeq\epsilon_H\equiv-\frac{\dot H}{H^2}
\end{equation}
In this paper, we use $\epsilon_{H}$ to define the end of the inflation ($\epsilon_H=1$) and the onset of radiation domination ($\epsilon_H=2$).

The specific form of the dissipation coefficient $\Upsilon$ depends on the microscopic physics underlying the WI model. This coefficient is generally a function of infalton $\phi$ and the radiation temperature $T$: $\Upsilon\equiv \Upsilon(\phi,T)$. For many WI models, the derived $\Upsilon(\phi,T)$ can be parameterized in the following form~\cite{WI5}:
\begin{equation}
    \Upsilon=C_{\Upsilon}T^c\phi^pM^{1-p-c}
\end{equation}
where $C_{\Upsilon}$ is a dimensionless constant, with  $c$ and $p$ being numerical powers and $M$ a mass scale in the model.
In this work, we focus on two representative and well-motivated WI models, which give the specific forms of the dissipation coefficients.

(i). The axion-like WI model introduces an axion-like coupling of the inflaton to a pure Yang-Mills gauge group, described by the following Lagrangian
\begin{equation}\label{sphL}
    \mathcal{L}_{\rm int}=\frac{\alpha_g}{8 \pi}\frac{\phi}{f_a}\tilde{F}^{a\mu\nu}F_{\mu\nu}^a
\end{equation}
where $F^a_{\mu\nu}$ is the field strength tensor and $\tilde{F}^{a\mu\nu}=\frac{1}{2}\epsilon^{\mu\nu\rho\sigma}F^a_{\rho\sigma}$. $\alpha_g \equiv g^2/4\pi$ with $g$ being the gauge coupling and $f_a$ corresponds to the axion decay constant in axion-like models.
This interaction triggers sphaleron processes in the thermal bath, leading to a dissipation coefficient of the form~\cite{Berghaus:2019whh,Laine:2021ego,Berghaus:2025dqi}.
\begin{equation}
    \Upsilon(T)=\frac{\Gamma_{\rm sph}(T)}{2f_a^2T}
\end{equation}
where $\Gamma_{\rm sph}(T)$ is the sphaleron rate which scales as $\Gamma_{\rm sph}(T) \sim \alpha_g^5T^4$~\cite{sph1}. So one can obtain $\Upsilon\propto T^3$.

(ii). The warm little inflation model~\cite{Bastero-Gil:2016qru} proposes a scenario which is similar to the little Higgs model. The inflaton is a pseudo-Nambu-Goldstone boson of a broken gauge symmetry. The dissipation coefficient is derived as $\Upsilon\propto T$ from the inflaton's coupling to light fermions.

We will discuss these two cases separately, $\Upsilon \propto T$, $T^3$, and compare their corresponding predictions. For each case, we consider $Q_{\rm ini}=0.01$ and $Q_{\rm ini}=1$, where $Q_{\rm ini}$ denotes the initial value of $Q$ at the beginning of inflation. Using the \texttt{WI2easy} code~\cite{WI2easy}, we have verified that both values satisfy the CMB constraints from \textit{Planck} data~\cite{Planck:2018jri}.

\section{Freeze-in production of gravitons in warm inflation}\label{FIPG}

In WI, gravitons are continuously produced from the plasma via particle scattering processes during and after inflation. Since gravitons are produced out of equilibrium and subject to Planck-scale suppression, their production mechanism is conceptually very similar to the ultraviolet (UV) DM freeze-in production mechanism~\cite{Ghiglieri:2022rfp}. And in ref.~\cite{Freese:2024ogj}, the authors found that freeze-in-produced DM relic density could be significantly enhanced in WI compared with CI (a brief review is provided in appendix~\ref{appA}). This conceptual similarity suggests that the freeze-in production of gravitons should be similarly enhanced in WI, which we focus on in this section.

In this work, we only consider gravitons with sufficiently high momenta to ensure that they always remain within the horizon. Their evolution is governed by the Boltzmann equation~\cite{Xu:2025wjq}
\begin{equation}\label{fh}
    \frac{\partial f_h}{\partial t}-kH\frac{\partial f_h}{\partial k}=C_h(k,T)
\end{equation}
where $f_h=f_h(t,k)$ and $k$ represent the distribution function of gravitons and graviton momentum, respectively. $C_h$ is the collision term for graviton production. For freeze-in process,  the collision term is a function of the plasma temperature and graviton momentum, but independent of $f_h$ since the gravitons can never reach equilibrium and the backreactions from gravitons can be safely neglected. 

The GW energy density is given by
\begin{equation}
    \rho_{\rm GW}=g_h\int\frac{d^3\boldsymbol{k}}{(2\pi)^3}kf_h
\end{equation}
with $g_h=2$ the degrees of freedom of the graviton.
 Now, by integrating eq.~\eqref{fh}, we get evolution equation for GW energy density:
\begin{equation}\label{evolution}
    \frac{d\rho_{\mathrm{GW}}}{dt}+4H\rho_{\mathrm{GW}}=\gamma
\end{equation}
where
\begin{equation}
    \gamma\equiv g_h\int\frac{d^3\boldsymbol{k}}{(2\pi)^3}kC_h(k,T)=g_h\int\frac{dk}{2\pi^2}k^3C_h(k,T)
\end{equation}
Now $\gamma$ corresponds to the graviton production rate density from the SM thermal bath and is a function of temperature $T$. Note that $\gamma$ has mass dimension 5. 
The leading-order contribution to $\gamma$ arises from single-graviton production processes (for example, the scattering process $ab\rightarrow ch$ where $a, b, c$ are SM particles and $h$ the graviton). The scattering amplitude for these processes is $\propto 1/M_{\rm pl}$, thereby suppressing the collision term—and hence $\gamma$ itself at leading order—by $1/M^2_{\rm pl}$. It then follows from dimensional analysis that the leading order term must scale as $T^7/M_{\rm pl}^2$. The next-to-leading order term, originating from double-graviton production~\cite{2graviton}, is further suppressed by $1/M_{\rm pl}^4$, and thus scales as $T^9/M_{\rm pl}^4$.
Therefore, via dimensional analysis, we obtain: 

\begin{equation}\label{gamma}
    \mathcal{\gamma}(T)\approx \mathcal{C}_1\frac{T^7}{M_{\rm pl}^2}+\mathcal{C}_2\frac{T^9}{M_{\rm pl}^4}+...
\end{equation}
\{where $\mathcal{C}_1$, $\mathcal{C}_2$ are coefficients which depend on the couplings of the thermal bath particles appearing in the collision term. For simplicity, we ignore the running of the couplings and assume they are constants. A detailed calculation of the collision term can be found in refs.~\cite{Ghiglieri:2020mhm,Ghiglieri:2022rfp,2graviton}. Here, we employ only the results of the dimensional analysis given in eq.~\eqref{gamma}, which suffices to provide preliminary insights into graviton production in WI. Moreover, since $T\ll M_{\rm pl}$ during WI as will be shown later, we consider only the leading-order contribution.

Using $dN\equiv Hdt$, where $N$ is the number of e-folds, the left-handed side of eq.~\eqref{evolution} can be re-parametrized as
\begin{equation}\label{re}
    \frac{d\rho_{\mathrm{GW}}}{dt}+4H\rho_{\mathrm{GW}} = He^{-4N} \frac{d(\rho_{\mathrm{GW}}e^{4N})}{dN}
\end{equation}
Then the solution of eq.~\eqref{re} reads
\begin{equation}\label{solution}
    \rho_{\rm GW}=e^{-4N}\int_{N_0}^{N}I_{\rm GW}(N')\ dN'
\end{equation}
where 
\begin{equation}\label{igw}
    I_{\rm GW}(N)\equiv e^{4N}\frac{\mathcal{\gamma}}{H}=\mathcal{C}_1 e^{4N}\frac{T^7(N)}{M_{\rm pl}^2H(N)}
\end{equation}
Note that $I_{\rm GW}$ is just the production rate of the comoving GW energy density, $e^{4N}\rho_{\rm GW}$, i.e. $I_\mathrm{GW}=d(e^{4N}\rho_\mathrm{GW})/dN$. And in deriving eq.~\eqref{solution}, we have assumed a vanishing GW energy density at the initial time.

From eq.~\eqref{igw}, during the inflation stage, we can see that  $I_\mathrm{GW}$ is an exponentially increasing function. This behavior arises because the inflaton energy dominates the Universe, as well as the temperature $T$ and the Hubble parameter $H$ remain approximately constant during the inflation stage. And when the Universe transits to the radiation domination stage, $I_{\mathrm{GW}}$ will be proportional to $e^{-N}$ since $T\propto e^{-N}$ and $H\propto e^{-2N}$. Therefore, $I_\mathrm{GW}(N)$ should sharply peaked at some e-fold, $N_{\mathrm{peak}}^{\mathrm{GW}}$, which satisfies $dI_{\rm GW}/dN=0$, or more explicitly:
\begin{equation}
     \left.\left[4+7\frac{d\ln{T(N)}}{dN}-\frac{d\ln{H(N)}}{dN}\right]\right|_{N = N_{\mathrm{peak}}^{\mathrm{GW}}}=0
\end{equation}
Thus, GW production is peaked around this maximum, and
the integral in eq.~\eqref{solution} is dominated by contributions from the vicinity of $N_{\rm peak}^{\rm GW}$. This allows us to estimate the GW energy density at some late times for $N>N_{\rm peak}^{\rm GW}$
\begin{equation}\label{GWestimate}
    \rho_{\rm GW}(N)\simeq e^{-4N}I_{\rm GW}(N_{\rm peak}^{\rm GW})\Delta N_{\mathrm{peak}}^{\rm GW}\quad(N>N_{\mathrm{peak}}^{\rm GW})
\end{equation}
with $\Delta N_{\mathrm{peak}}^{\rm GW}\gtrsim 1$ being the half-width of $I_{\rm GW}(N)$.

\begin{figure}[htbp]
	\centering
	\begin{minipage}{0.5\linewidth}
		\centering
		\includegraphics[width=1.\linewidth]{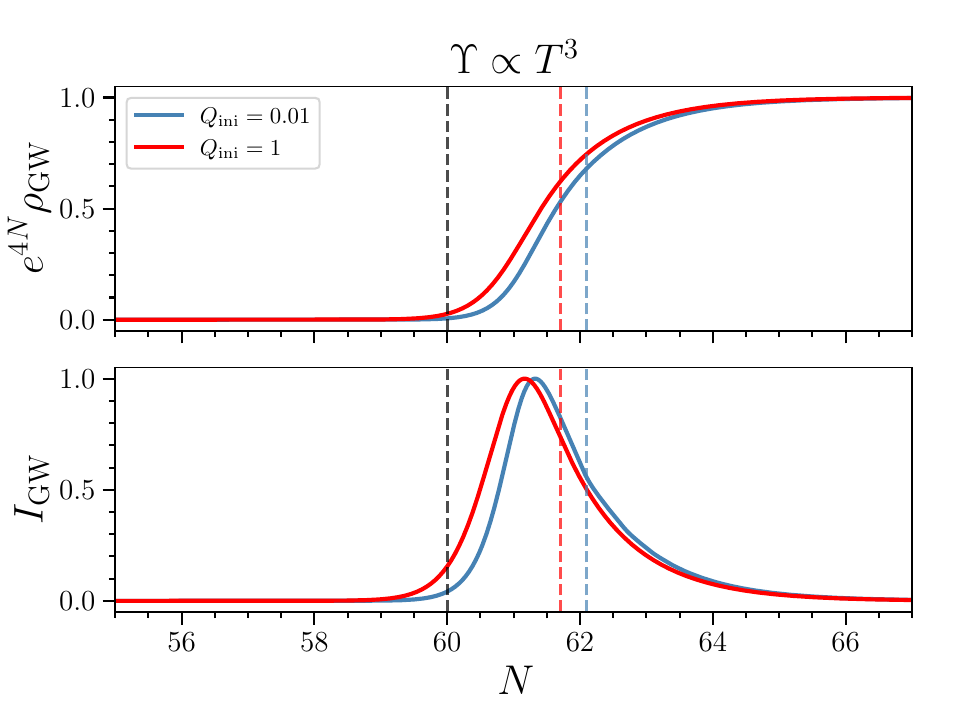}
	\end{minipage}%
	\begin{minipage}{0.5\linewidth}
		\centering
		\includegraphics[width=1.\linewidth]{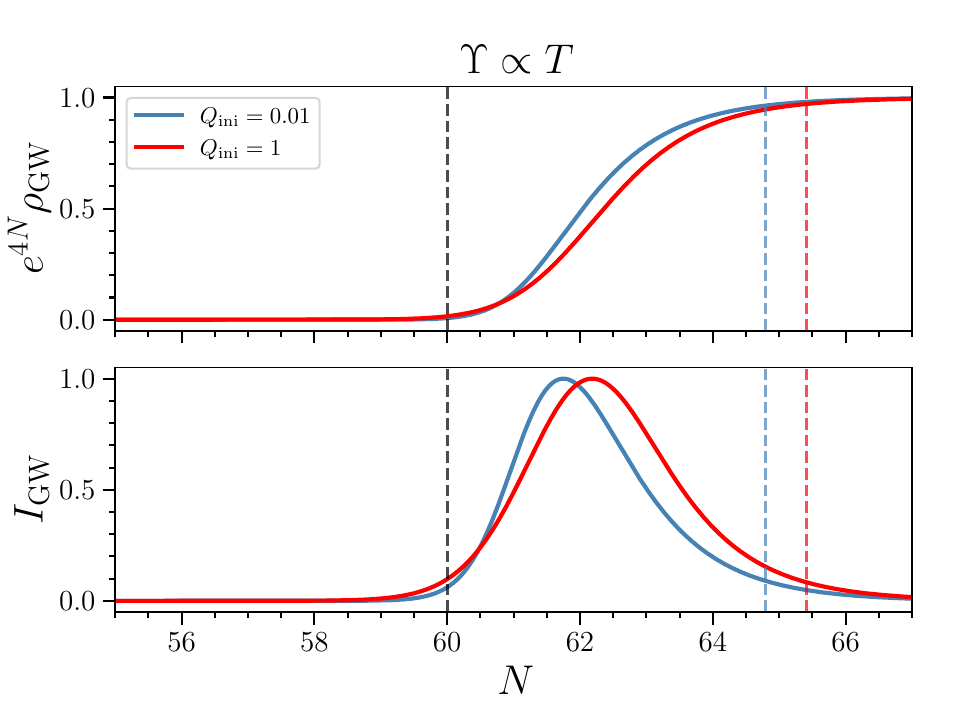}
	\end{minipage}
    \caption{The comoving GW energy density $e^{4N}\rho_{\rm GW}$ and its production rate $I_{\rm GW}$ in WI with dissipation term $\Upsilon\propto T^3$ and $\Upsilon\propto T$, respectively. All curves are normalized to their respective maximum values. The blue line corresponds to $Q_{\mathrm{ini}}=0.01$ and the red line corresponds to $Q_{\mathrm{ini}}=1$. The black dashed line indicates the end of inflation ($\epsilon_{H}=1$), while the red and blue dashed lines correspond to the onset of radiation domination ($\epsilon_{H}=2$), respectively. Note that the reheating process is more efficient for $\Upsilon\propto T^3$ than for $\Upsilon\propto T$.}
    \label{fig:GWdensity}
\end{figure}

The evolution of $I_{\rm GW}$ and $e^{4N}\rho_{\rm GW}$ are presented in figure~\ref{fig:GWdensity}. It can be clearly seen that most GW are produced around the peak of $I_{\rm GW}$, which occurs after the end of inflation but before the radiation-dominated epoch. 

To better understand the freeze-in production of gravitons in WI, we compare it to the CI scenario. In CI, the production of freeze-in gravitons commences only after the Universe enters the radiation-dominated era.
In contrast, gravitons can be produced over the entire inflationary epoch in WI, and as we have seen that most graviton energy is produced before the onset of radiation domination ($\epsilon_{H}=2$). Thus, we expect graviton production to be more efficient in WI. It is therefore natural to quantify the enhanced production in WI by comparing the total GW energy density in WI with that produced solely during the radiation-dominated epoch. Equivalently, this corresponds to comparing the gravitons production in WI with that in a CI scenario having the same reheating temperature $T_{\rm RH}$, which is defined as  $T_{\rm RH}\equiv T(N_{\rm RH})$ with $N_{\rm RH}\equiv N(\epsilon_{H}=2) $.

It is then useful to normalize the GW energy density by defining the dimensionless quantity $\rho_{\rm GW}/s^{\frac{4}{3}} $ with $s=\frac{2\pi^2g_{*s}}{45}T^3$ the entropy density of the thermal bath~\cite{Ghiglieri:2015nfa}. Eq.~\eqref{GWestimate} gives:
\begin{equation}
    \frac{\rho_{\rm GW}}{s^{\frac{4}{3}}}\simeq \left(\frac{2\pi^{2}g_{*s}}{45}\right)^{-\frac{4}{3}}\mathcal{C}_1\frac{e^{4(N_{\mathrm{peak}}^{\rm GW}-N)}}{M_{\rm pl}^2T^{4}(N)}\Delta N_{\mathrm{peak}}^{\rm GW}
\times\frac{ T^{7}(N_{\mathrm{peak}}^{\rm GW})}{H(N_{\mathrm{peak}}^{\rm GW})}\quad(N>N_{\mathrm{peak}}^{\rm GW})
\end{equation}

For CI with the same reheating temperature $T_{\mathrm{RH}}$, by integrating eq.~\eqref{solution} from $N_{\mathrm{RH}}$ to infinity we can obtain
\begin{equation}
   \bigg[ \frac{\rho_{\rm GW}}{s^{\frac{4}{3}}}\bigg]_{\mathrm{CI}} \simeq \left(\frac{2\pi^2g_{*s}}{45}\right)^{-\frac{4}{3}}\mathcal{C}_1\frac{T_{\rm RH}^3}{M_{\rm pl}^2H_{\rm RH}}
\end{equation}
where $H_{\rm RH}$ is the Hubble parameter at $N_{\rm RH}$.

The GW production in WI is therefore enhanced by a factor of 
\begin{equation}
  \bigg[ \frac{\rho_{\rm GW}}{s^{\frac{4}{3}}}\bigg]\bigg/\bigg[ \frac{\rho_{\rm GW}}{s^{\frac{4}{3}}}\bigg]_{\mathrm{CI}}\simeq \frac{I_{\rm GW}(N_{\rm peak}^{\rm GW})}{I_{\rm GW}(N_{\rm RH})}\Delta N_{\mathrm{peak}}^{\rm GW}
\end{equation}

We conclude that GW are primarily generated around $N_{\rm peak}^{\rm GW}$, with their production enhanced in WI. Moreover, we expect a more pronounced enhancement for the WI model with $\Upsilon\propto T$ than for the $\Upsilon\propto T^3$ model, as the more efficient reheating process in the latter case leads to a significantly smaller value of  the ratio $I_{\rm GW}(N^{\rm GW}_{\rm peak})\big/I_{\rm GW}(N_{\rm RH})$.

\section{Gravitational waves spectrum}\label{SPECTRA}

Now we turn to the analysis of the GW spectrum. The graviton production rate density $\gamma$ takes the form~\cite{Ghiglieri:2015nfa,Ghiglieri:2020mhm}
\begin{equation}
    \gamma(T)=\frac{4T^7}{M^2_{\rm pl}}\int\frac{d^3\hat{k}}{(2\pi)^3}\hat{\eta}(\hat{k},T)
\end{equation}
with$\ \hat{k}=k/T$. The $\hat{\eta}$ function takes into account all the Standard Model contributions and has been calculated in refs.~\cite{Ghiglieri:2015nfa,Ghiglieri:2020mhm} using the leading-log approximation. The calculation including hard thermal loop resummation has been done in ref.~\cite{Ringwald:2020ist}. Here we only consider the results of  leading-log approximation~\cite{Ghiglieri:2015nfa}, 
\begin{equation}
    \hat{\eta}(\hat{k},T)=\frac{\hat{k}}{e^{\hat{k}}-1}\sum_{i}d_i\hat{m}_i^2(T)\ln\left(\frac{5}{\hat{m}_i(T)}\right)
\end{equation}
with 
\begin{equation}
\begin{aligned}
d_1 &=1, \quad d_2 =3, \quad d_3=8,\\
\hat{m}_1^2 &= \frac{11}{6}g_1^2, \quad \hat{m}_2^2 = \frac{11}{6}g_2^2, \quad \hat{m}_3^2 = g_3^2
\end{aligned}
\end{equation}
where $g_1$, $g_2$, $g_3$ are gauge couplings of the Standard Model $U(1)_Y$, $SU(2)_L$ and $SU(3)_C$ gauge groups, respectively.

The GW energy density is given by eq.~\eqref{solution} as
\begin{equation}
    \rho_{\mathrm{GW}}(N)=\frac{2}{\pi^2M_{\rm pl}^2}e^{-4N}\int_{N_0}^{N}dN' \ e^{4N'}\frac{T^7}{H}\int d\hat{k}\  \hat{k}^2\hat{\eta}(\hat{k},T)
\end{equation}
where $N$ is chosen in the radiation-dominated era and long after the peak of $I_{\mathrm{GW}}$.

We can then calculate the GW energy spectrum
\begin{equation}
    \Omega_{\mathrm{GW}}=\frac{1}{\rho_c}\frac{d\rho_{\mathrm{GW}}}{d\ln k}
\end{equation}
where $\rho_c$ is the critical density of Universe. We redshift the GW spectrum to the present time using
\begin{equation}
    \rho_{\gamma}=\rho_{\gamma,0}\left(\frac{a_0}{a}\right)^4\left(\frac{g_{*s,0}}{g_{*s}(N)}\right)^{\frac{4}{3}}\quad,\quad \frac{k(N)}{T(N)}=\frac{k_0}{T_0}\left(\frac{g_{*s}(N)}{g_{*s,0}}\right)^{\frac{1}{3}}
\end{equation}
where $g_{* s, 0}= 3.931$~\cite{Planck:2018vyg} is the present degrees of freedom of the entropy density. $\rho_{\gamma,0}$ is the current photon energy density.
\begin{figure}
	\centering
	\begin{minipage}{0.8\linewidth}
		\centering
		\includegraphics[width=1.\linewidth]{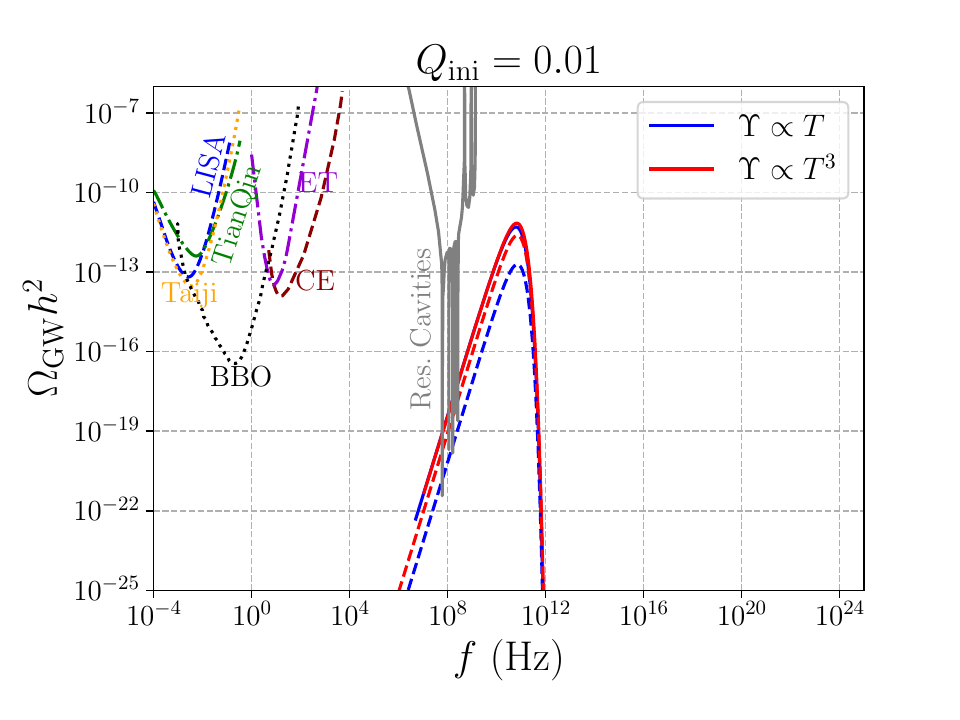}
	\end{minipage}
	\begin{minipage}{0.8\linewidth}
		\centering
		\includegraphics[width=1.\linewidth]{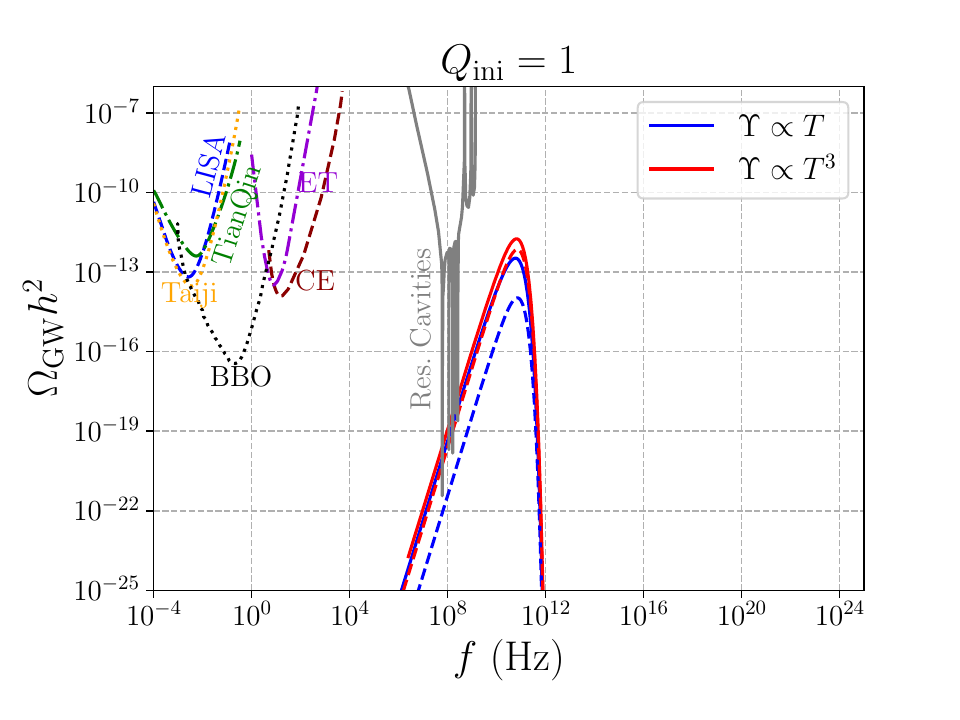}
	\end{minipage}
	\qquad

		\label{chutian3}
    \caption{The freeze-in GW spectrum of WI in the high frequency regime for different $Q_{\rm ini}$ and different dissipation terms. The dashed lines correspond to the freeze-in GW spectrum in CI produced during the radiation-dominated epoch with the same reheating temperature. The quartic coupling $\lambda$ is chosen to satisfy the CMB constraints. The upper panel and lower panel shows the GW spectrum for $Q_{\rm ini} = 0.01$ and $Q_{\rm ini} = 1$, respectively. Note the nearly identical spectral shapes and approximately equal peak frequencies, along with an enhanced peak amplitude compared to the CI case.}
    \label{fig:GW3}              
\end{figure}

The GW spectrum today is given by
\begin{equation}
\begin{split}
    \Omega_{\mathrm{GW},0}(k_0)h^2=&\frac{2}{\pi^2M_{\rm pl}^2}\frac{\Omega_{\gamma,0}h^2}{\rho_c(N)}\frac{g_{*}(N)}{g_\gamma}\left(\frac{g_{*s,0}}{g_{*s}(N)}\right)^{\frac{1}{3}}\times\\&e^{-4N}\int_{N_0}^{N}dN'\ e^{4N'}\frac{T^7}{H}\left(\alpha(N')\frac{k_0}{T_0}\right)^3\hat{\eta}\left[\alpha(N')\frac{k_0}{T_0}\left(\frac{g_{*s}}{g_{*s,0}}\right)^{\frac{1}{3}},T\right]
\end{split}
\end{equation}
where $\Omega_{\gamma,0} = 5.38 \times 10^{-5}$ , $g_\gamma = 2$ and the dimensionless Hubble parameter $h=0.674$~\cite{Planck:2018vyg}. We set $g_{*s}(N)= g_{*}(N) = 106.75$. And we have defined
\begin{equation}
    \alpha(N')\equiv\frac{T(N)e^{N}}{T(N')e^{N'}}
\end{equation}
Note that this function stays constant during the radiation-dominated era since $T(N')\propto e^{-N'}$. So in CI, as the GW are produced during the radiation-dominated era, the GW spectrum can be parametrized in the form of
\begin{equation}
    \Omega_{\mathrm{GW},0}^{\mathrm{CI}}(f)=\mathcal{A}\frac{(f/f_{\rm peak})^4}{e^{f/f_{\rm peak}}-1}
\end{equation}
where $f=\frac{k_0}{2\pi}$ is the present frequency. And the peak frequency can be derived from the maximum point of $\hat{k}^4/(e^{\hat{k}}-1)$,
\begin{equation}
    f_{\rm peak}^{\rm CI} \simeq 3.92 \frac{T_0}{2\pi} \left(\frac{g_{*s,0}}{g_{*s}}\right) ^{\frac{1}{3}}\approx7.41 \times 10^{10}~\mathrm{Hz}
\end{equation}
In WI, most GWs are generated around $N_{\mathrm{peak}}^{\mathrm{GW}}$, occurring after inflation and before the radiation-dominated era, as we have discussed. During this stage, $\alpha(N')$ will slowly evolve over time, so the shape of GW spectrum will differ from  $\Omega_{\mathrm{GW},0}^{\rm CI}$, and the peak frequency shifts. However,  as we will show in our result, this change is very slight and  $f_{\rm peak}^{\rm WI}\simeq f_{\rm peak}^{\rm CI}$, where $f_{\rm peak}^{\rm WI}$ is the peak frequency of freeze-in GW in WI.

In figure~\ref{fig:GW3}, we show the GW spectrum for different values of $Q_{\mathrm{ini}}$, both in the cases of $\Upsilon \propto T$ and $\Upsilon \propto T^3$. Additionally, the spectrum of freeze-in GW produced during the radiation-dominated epoch in CI are represented by dashed lines. The shapes of the spectra and the peak frequencies are nearly identical, while the peak amplitude is enhanced in WI. Note also that, as expected, this enhancement is much more pronounced in the $\Upsilon \propto T$ model than in the $\Upsilon \propto T^3$ case. We next delve into a comparison of the two models. When $Q_{\rm ini}=0.01$, the predictions of the two dissipative models are nearly indistinguishable. However, for  $Q_{\rm ini}=1$, differences arising from the temperature dependence of the dissipation coefficients lead to a noticeable divergence: the peak amplitude is lower in the $\Upsilon\propto T$ model than in the $\Upsilon\propto T^3$ model. We also plot the sensitivity curves of LISA~\cite{LISA:2017pwj}, TianQin~\cite{TianQin:2015yph}, Taiji~\cite{Hu:2017mde}, BBO~\cite{Crowder:2005nr,Corbin:2005ny}, Cosmic Explorer (CE)~\cite{Reitze:2019iox}, Einstein telescope (ET)~\cite{Sathyaprakash:2012jk,ET:2019dnz}, and Resonant Cavities~\cite{Herman:2022fau} for all panels.
\section{Freeze-in production of purely gravitational dark matter}\label{FIPPGDM}

Purely gravitational DM~\cite{GDMTang:2016,Garny:2017kha,GDMEma:2018ucl} is one of the simplest possibilities for DM, since gravitational interaction is universal and model-independent (at least for sub-Planckian scales). It can be produced by the thermal bath mediated by gravitons, known as the gravitational DM freeze-in and it is a concrete realization of the DM UV freeze-in mechanism. The production of purely gravitational DM in early Universe was carefully studied in CI~\cite{Tang:2017hvq,Bernal:2018qlk,GDMinCIHashiba:2018tbu,GDMinCIMambrini:2021zpp,Clery:2021bwz}, and was first considered in WI in ref.~\cite{wqy2025}. In this section, we focus on the correlation between the DM mass and the peak amplitude of the GW spectra as functions of the quartic coupling of inflaton field, applying our results from the last section.

We start with the Boltzmann equation of purely gravitational DM $\chi$ .
\begin{equation}\label{DMevolution}
    \frac{dn_{\chi}}{dt}+3H n_{\chi} =\langle\sigma v\rangle_{\chi}\big(n_\chi^{\rm eq}\big)^2
\end{equation}
where $n_{\chi}$ denotes the number density of Fermion DM $\chi$. And $n_\chi^{\mathrm{eq}}=\frac{3}{2}\frac{\zeta_3}{\pi^2}T^3$ is the DM number density in thermal equilibrium, where $\zeta_3=1.2\ $ is the value of Riemann-Zeta function.

The thermal averaged cross sections of all gravitational annihilation channels have been calculated in ref.~\cite{Garny:2017kha}
\begin{equation}
    \langle\sigma v\rangle_{0\rightarrow\chi}=\frac{\pi m_\chi T}{2M_{\rm pl}^4}\left[\frac{4}{5}\frac{T}{m_\chi}+\frac{1}{5}\frac{m_\chi}{T}-\frac{1}{5}\frac{m_\chi}{T}\frac{K_1^2}{K_2^2}+\frac{2}{5}\frac{K_1}{K_2}   \right]
\end{equation}
\begin{equation}
    \langle\sigma v\rangle_{\frac{1}{2}\rightarrow\chi}=\langle\sigma v\rangle_{1\rightarrow\chi}=\frac{4\pi m_{\chi}T}{M_{pl}^4}\left[\frac{6}{5}\frac{T}{m_\chi}+\frac{2}{15}\frac{m_\chi}{T}-\frac{2}{15}\frac{m_\chi}{T}\frac{K_1^2}{K_2^2}+\frac{3}{5}\frac{K_1}{K_2}  \right]
\end{equation}
where$\ K_i=K_i(m_\chi/T)$ are the modified Bessel functions.
In the limit of $m_\chi \ll T$, we have
\begin{equation}
    \langle\sigma v\rangle_{0\rightarrow\chi}=\frac{2\pi T^2}{5 M_{\rm pl}^4}\quad , \quad\langle\sigma v\rangle_{\frac{1}{2}\rightarrow\chi}=\langle\sigma v\rangle_{1\rightarrow\chi}=\frac{24\pi T^2}{5 M_{\rm pl}^4}
\end{equation}

Thus the total cross section gives
\begin{equation}
    \langle\sigma v\rangle_\chi=4\langle\sigma v\rangle_{0\rightarrow\chi}  +45 \langle\sigma v\rangle_{\frac{1}{2}\rightarrow\chi}+ 12\langle\sigma v\rangle_{1\rightarrow\chi}=\frac{1376\pi T^2}{5 M_{\rm pl}^4}
\end{equation}
\begin{figure}[h]
	\centering
	\begin{minipage}{0.5\linewidth}
		\centering
		\includegraphics[width=1.0\linewidth]{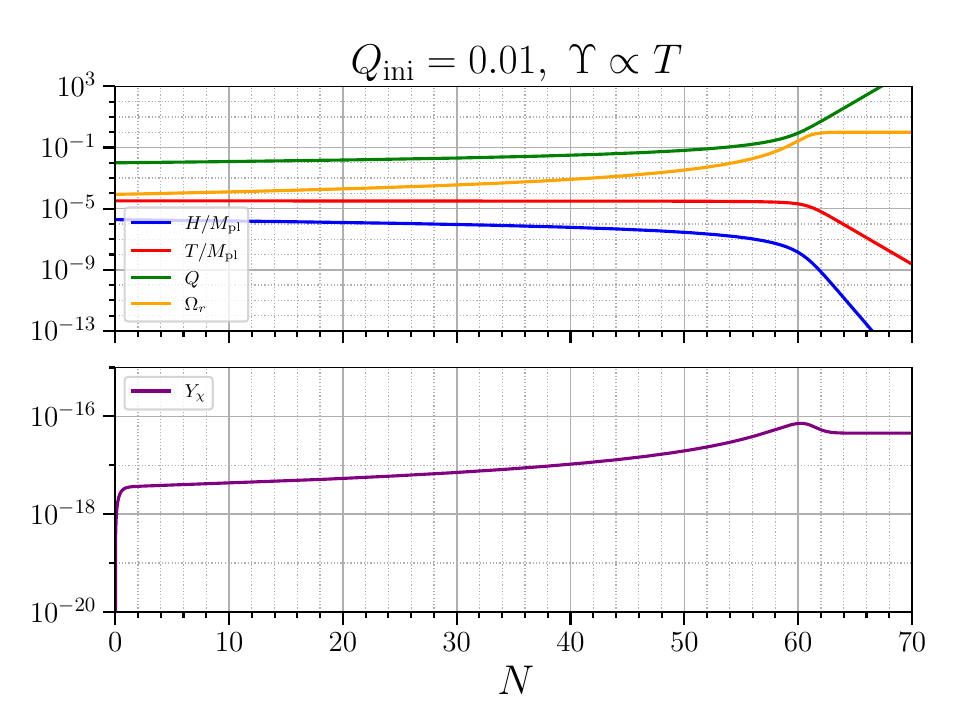}
	\end{minipage}%
\hfill
	\begin{minipage}{0.5\linewidth}
		\centering
		\includegraphics[width=1.0\linewidth]{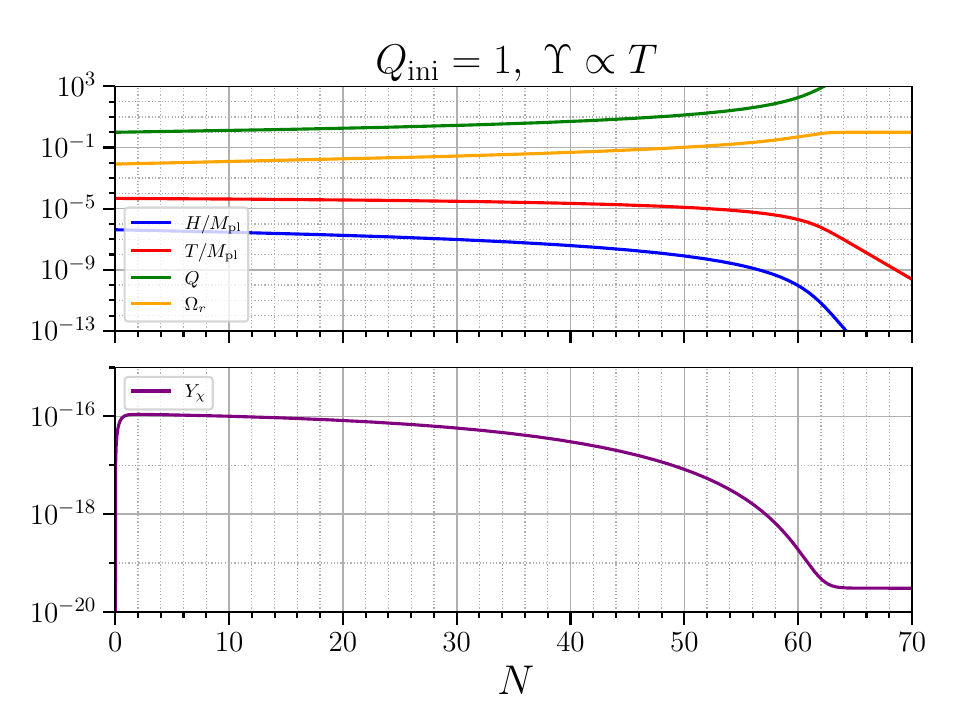}
	\end{minipage}
    \begin{minipage}{0.5\linewidth}
        \centering
        \includegraphics[width=1.0\linewidth]{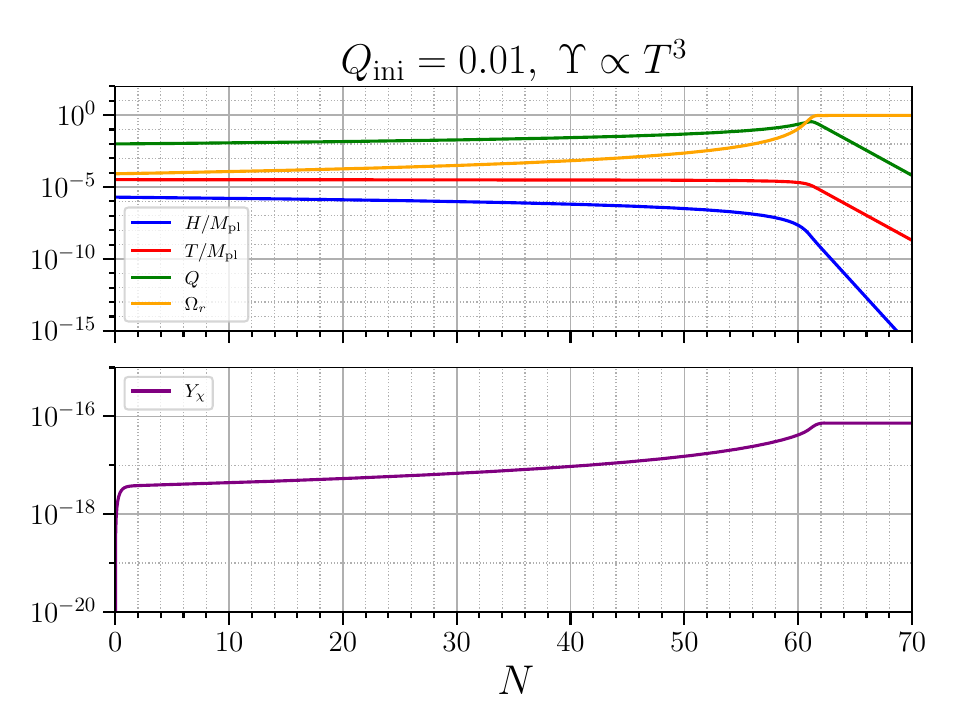}
    \end{minipage}%
    \begin{minipage}{0.5\linewidth}
		\centering
		\includegraphics[width=1.0\linewidth]{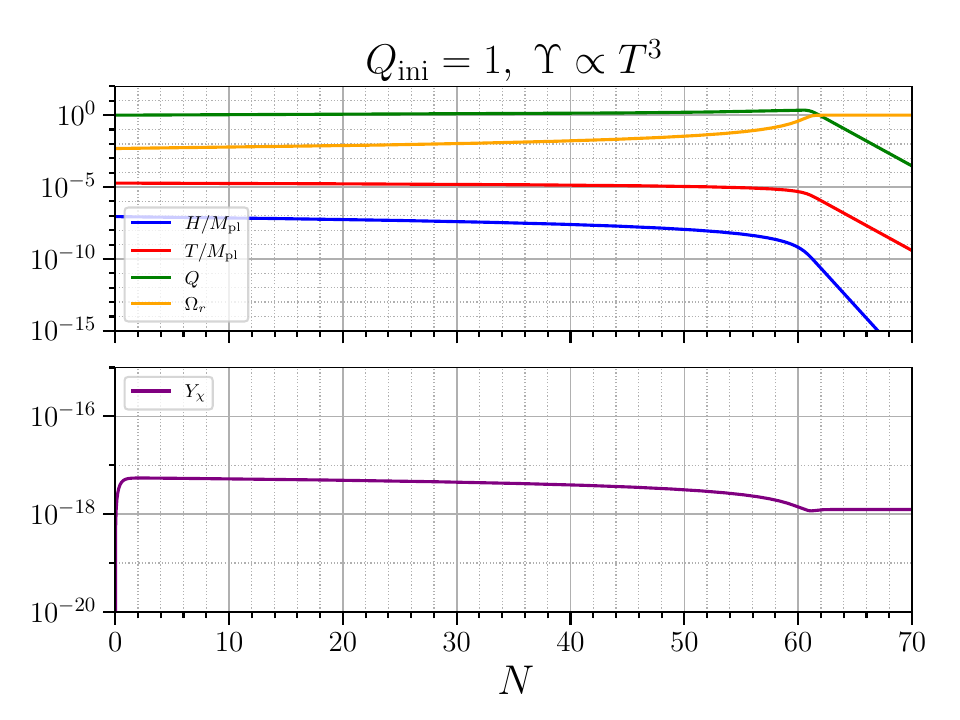}
	\end{minipage}
        \caption{The evolution of the Hubble parameter $H$, the plasma temperature $T$, the parameter $Q$, the radiation density parameter $\Omega_r$ and the DM yield $Y_\chi$.}
        \label{fig:evolution}
\end{figure}

Now eq.~\eqref{DMevolution} becomes
\begin{equation}
    He^{-3N}\frac{d n_\chi e^{3N}}{dN}=\beta\frac{T^8}{M_{\rm pl}^4}
\end{equation}
where $\beta=3096\zeta_3^2/5\pi^3$. Then the solution of DM number density can be expressed as
\begin{equation}
    n_\chi(N)=e^{-3N}\int_{N_0}^{N} I_{\chi}(N')dN' 
\end{equation}
where 
\begin{equation}
    I_{\chi}(N)= e^{3N}\beta\frac{T^8(N)}{M_{\rm pl}^4H(N)}
\end{equation}
and the DM yield $Y_\chi \equiv n_\chi/s$ reads
\begin{equation}\label{Ychi}
    Y_\chi(N)=\frac{45}{2\pi^2 g_{*s}}\frac{e^{-3N}}{T^3(N)}\int_{N_0}^{N} I_{\chi}(N')   dN'
\end{equation}

We show the evolution of $Y_\chi$ in figure~\ref{fig:evolution}. The evolution of the Hubble parameter $H$, the plasma temperature $T$ and the parameter $Q$ are also given in figure~\ref{fig:evolution}. During the early stage of WI, although the inflaton energy dominates the Universe, the thermal bath is still sustained with a nearly constant temperature and DM gradually accumulates. After the inflation stage, the Universe transits to a radiation-dominated phase, where $T \propto e^{-N}$ and $H \propto T^2$.

Assuming that DM comprises only $\chi$ particles, their mass can be derived from
\begin{equation}
    \Omega_{\mathrm{DM},0}h^2=\frac{m_\chi Y_\chi(N) s_0}{\rho_{c,0}}h^2 \equiv 0.12
\end{equation}
where $ \ s_0\simeq 2891.2 ~\mathrm{cm}^{-3} \ $ and $\rho_{c,0}\simeq 4.78\times 10^{-6} ~\mathrm{GeV} \cdot \mathrm{cm}^{-3}$~\cite{Planck:2018vyg} are the entropy density and the critical energy density today, respectively.

\begin{figure}[h]
	\centering
	\begin{minipage}{0.49\linewidth}
		\centering
		\includegraphics[width=1.0\linewidth]{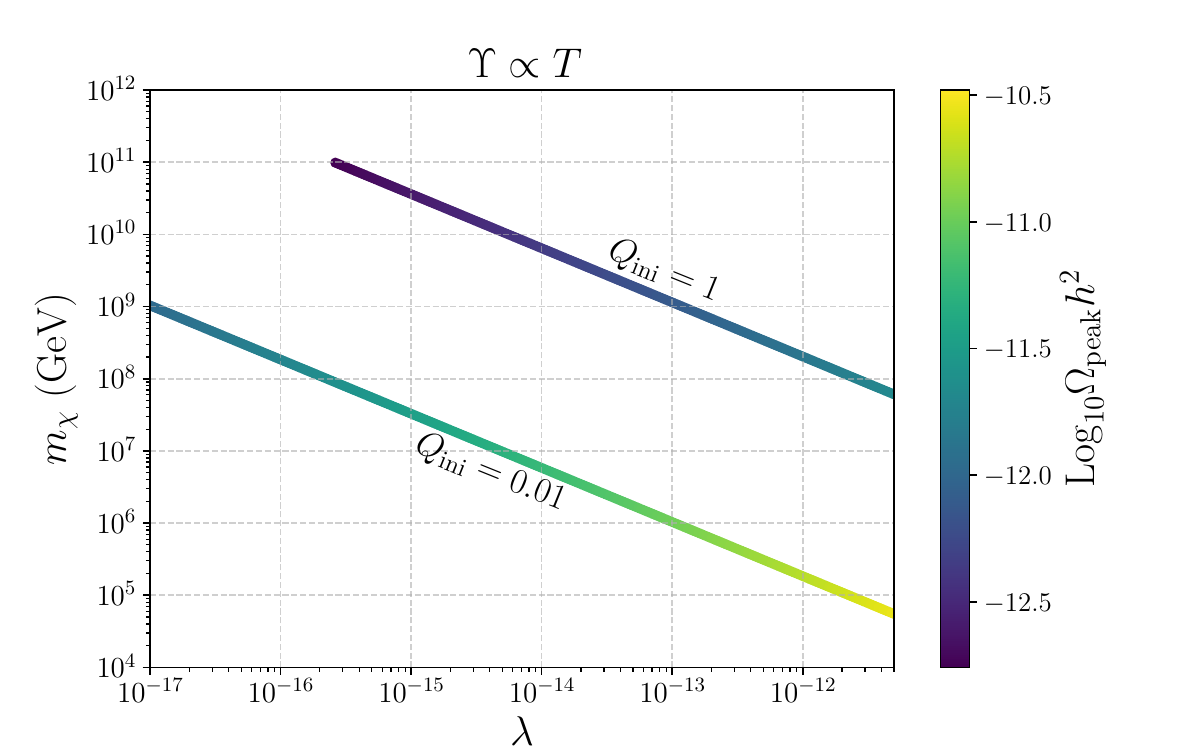}
	\end{minipage}
	\begin{minipage}{0.49\linewidth}
		\centering
		\includegraphics[width=1.0\linewidth]{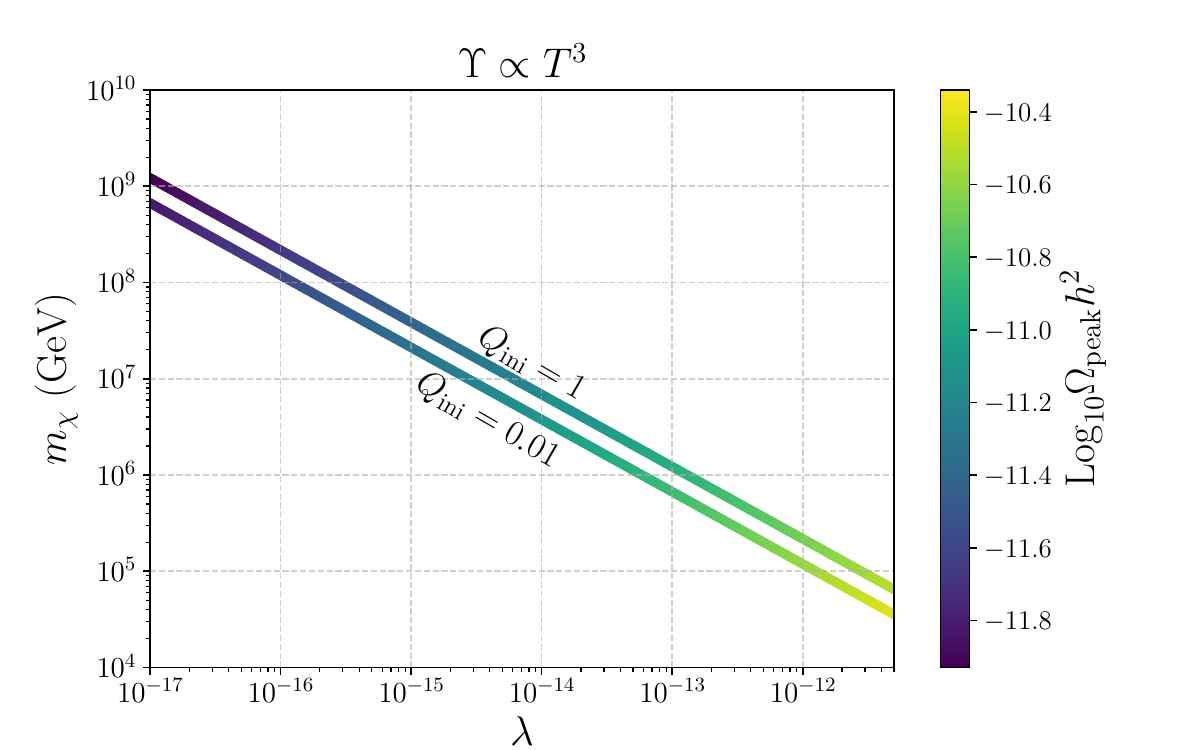}
	\end{minipage}
        \caption{The correlation between DM mass and the peak amplitude of GW spectra for different $Q_{\rm ini}$ and dissipation terms. The colored bar represents the peak amplitude of GW spectra fixing the observed DM relic density. In the upper panel where $\Upsilon \propto T$, we found that in the case of $Q_{\mathrm{ini}} = 1$, the DM mass does not satisfy the condition $m_\chi \ll T$ for $\lambda < 2.6\times 10^{-16}$.}
        \label{fig:GW_DM}
\end{figure}
\begin{figure}[h]
	\centering
	\begin{minipage}{0.49\linewidth}
		\centering
		\includegraphics[width=1.0\linewidth]{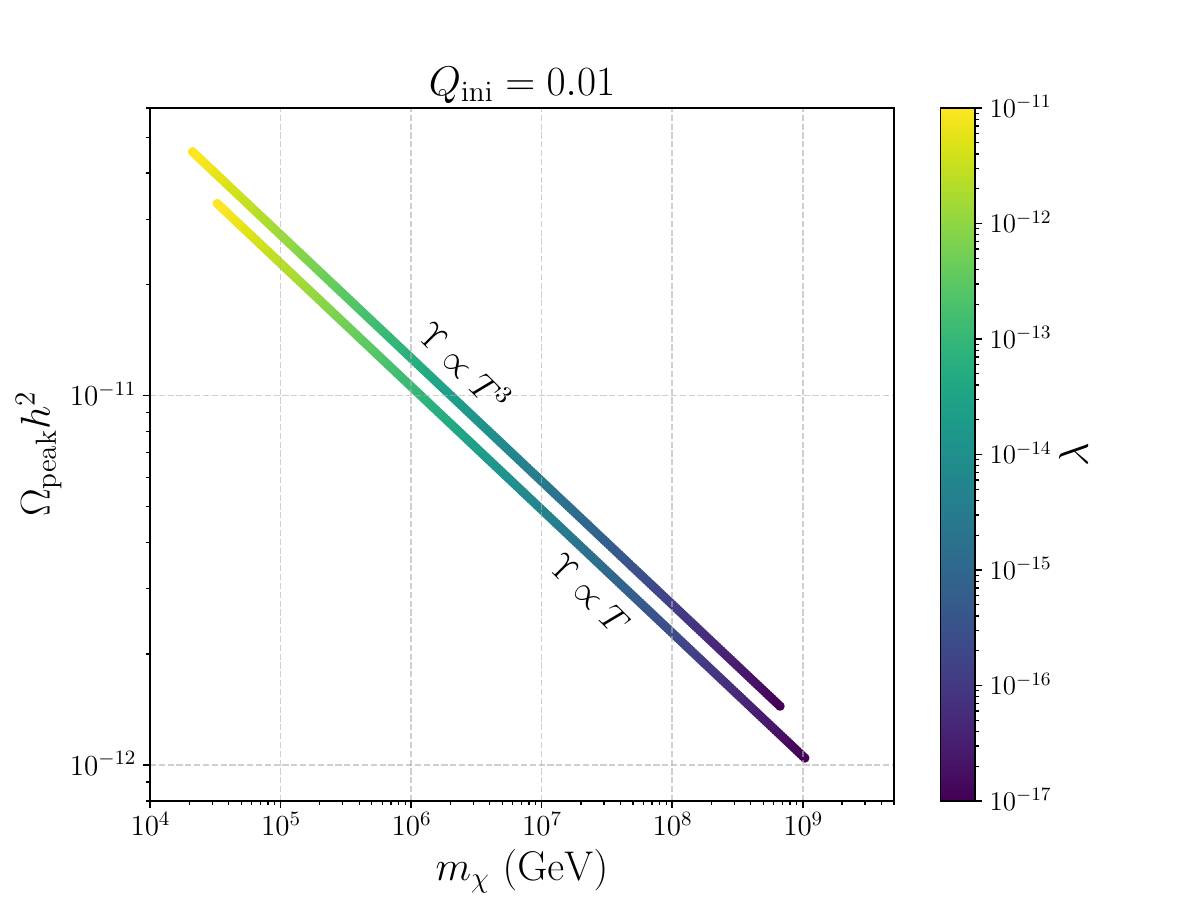}
	\end{minipage}
	\begin{minipage}{0.49\linewidth}
		\centering
		\includegraphics[width=1.0\linewidth]{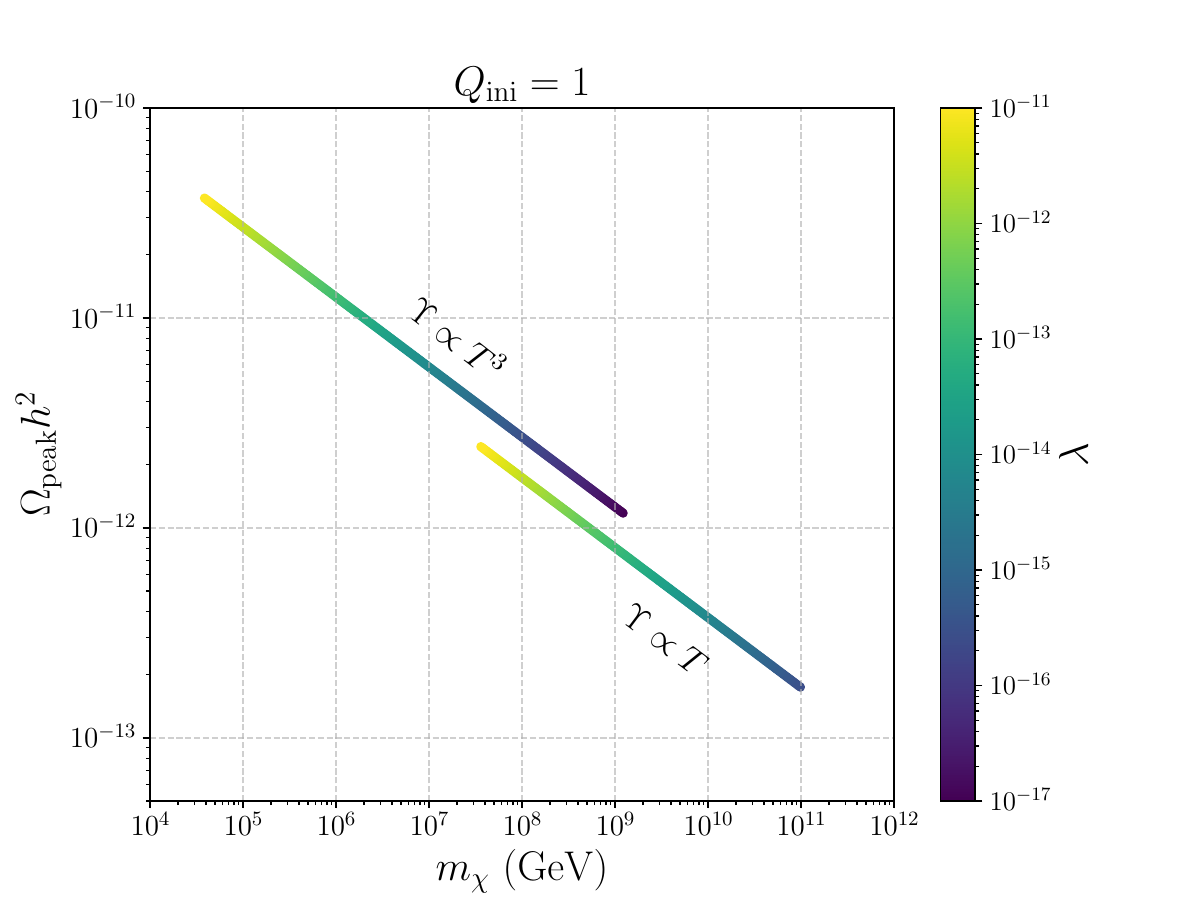}
	\end{minipage}
        \caption{The correlation between DM mass and the peak amplitude of GW spectra for different $Q_{\rm ini}$ and dissipation terms, with $\Omega_{\rm peak}h^2$ on the y-axis, $m_\chi$ on the x-axis, and $\lambda$ color-coded.}
        \label{fig:new}
\end{figure}

In figure~\ref{fig:GW_DM} and figure~\ref{fig:new}, to satisfy the observed DM relic density, we show the correlation between DM mass and the peak amplitude of GW spectra for different $Q_{\rm ini}$ and dissipation terms. For each point on the curves, we have checked that the condition $m_\chi \ll T$ is always satisfied at $N=N_{\rm peak}^{\rm DM}$, when most DM is produced. Compared with the case $\Upsilon \propto T^3$, it can be seen that varying $Q_{\rm ini}$ has a stronger impact on both the predicted DM mass and the peak amplitude of the GW spectra in the case of $\Upsilon\propto T$. Moreover, the DM mass is significantly more sensitive to the coupling $\lambda$ than the GW peak amplitude is. And as clearly shown in figure~\ref{fig:new}, the peak amplitude of the GW spectra decreases with increasing DM mass across all cases studied. Notably, the distinction between the two WI models for $Q_{\rm ini}=1$ becomes more pronounced than that for $Q_{\rm ini}=0.01$. 

For a given $Q_{\rm ini}$, the value of $\lambda$ can be fixed by matching the observed scalar spectrum amplitude $A_s\approx2.1\times10^{-9}$~\cite{Planck:2018jri,WI2easy}. Once $\lambda$ is determined, we then obtain the corresponding DM mass as well as the peak amplitude of the GW spectrum. Table~\ref{tab:placeholder1} and table~\ref{tab:placeholder2} list the values of $\lambda$, the scalar spectral index $n_s$, the tensor-to-scalar ratio $r$, DM mass, peak frequency, and the peak amplitude of the GW signal for WI model with $\Upsilon\propto T$ and $\Upsilon\propto T^3$, respectively. Notably, the GW spectra show nearly identical peak frequencies with amplitudes of order $10^{-13}-10^{-12}$, consistent with our previous results. The DM mass, however, varies significantly across models, ranging from $10^6$ to $10^{10}$ GeV. 
\vspace{1em}
\begin{table}[h]
    \centering

    \begin{tabular}{c|c|c|c|c|c|c}\hline\hline
         $Q_{\mathrm{ini}}$&   $\lambda$ & $n_s$  & $r$ & $f_{\mathrm{peak}}(\mathrm{Hz})$& $\Omega_{\mathrm{peak}}h^2$&$m_{\chi}(\mathrm{GeV})$\\\hline
         0.01&  $ 5.26\times10^{-15}$  &$0.964$  &   $8.05\times10^{-3}$ & $6.31\times10^{10}$& $ 5.00\times10^{-12}$&$9.44\times10^{6}$\\\hline
 1&  $3.55\times10^{-15}$  & $0.969$  & $3.75\times10^{-4}$ & $6.03\times10^{10}$& $3.34\times10^{-13}$&$1.38\times10^{10}$\\\hline\end{tabular}
\caption{Model predictions for WI with $\Upsilon\propto T$: scalar spectral index $n_s$, tensor-to-scalar ratio $r$, peak frequency $f_{\mathrm{peak}}$, peak amplitude $\Omega_{\mathrm{peak}}h^2$, and DM mass $m_\chi$.}
\label{tab:placeholder1}
    
\end{table}
\begin{table}[h]
    \centering

    \begin{tabular}{c|c|c|c|c|c|c}\hline\hline
         $Q_{\mathrm{ini}}$&   $\lambda$ & $n_s$  & $r$ & $f_{\mathrm{peak}}(\mathrm{Hz})$& $\Omega_{\mathrm{peak}}h^2$&$m_{\chi}(\mathrm{GeV})$\\\hline
  0.01&  $ 5.52\times10^{-15}$&$0.963$  &   $8.87\times10^{-3}$ & $6.61\times10^{10}$& $6.99\times10^{-12}$&$5.88\times10^{6}$\\\hline
 1&  $5.41\times10^{-17}$& $0.969$  & $1.63\times10^{-5}$
  & $6.61\times10^{10}$& $1.79\times10^{-12}$&$3.44\times10^{8}$\\ \hline\end{tabular}
\caption{Model predictions for WI with $\Upsilon\propto T^3$: scalar spectral index $n_s$, tensor-to-scalar ratio $r$, peak frequency $f_{\mathrm{peak}}$, peak amplitude $\Omega_{\mathrm{peak}}h^2$, and DM mass $m_\chi$.}
\label{tab:placeholder2}
    
\end{table}

\section{Conclusions}\label{Conculsions}

Motivated by the recent development in WI model~\cite{Berghaus:2025dqi}, we investigate two important cosmic relics: GW and DM, in the WI scenarios. We have extended the framework of freeze-in graviton production from CI to WI, computing the resulting GW spectrum today in two well-motivated WI models, characterized by dissipation coefficients $\Upsilon\propto T$ and $\Upsilon\propto T^3$. We find that the freeze-in production of GW is enhanced in WI, similar to the case of UV freeze-in DM studied in previous literature.
Moreover, the shapes of GW spectra and  the peak frequencies ($\sim10^{10}~\rm Hz$) are nearly identical across different WI models, while the peak amplitude of the GW spectra today can be of the order $10^{-13}-10^{-12}$. We also studied graviton-portal DM within WI framework, as a specific and simple possibility of UV freeze-in DM.
Under the constraint of reproducing the observed DM relic abundance, we have identified a correlation between the DM mass and the quartic coupling of the inflation field, and determined the corresponding peak amplitude of the GW signal. Notably, the predicted DM mass exhibits significant variation across different models and parameters.

In summary, our work explores the connections between GWs, DM, and the microscopic dynamics of WI. We observe that different WI models, characterized by different dissipation coefficients and hence different background evolution, lead to distinguishable GW signals and DM relic abundances. This suggests that future observations combining high-frequency GW detection and DM phenomenology could offer a potential approach to probing the detailed physics of WI models.

\begin{acknowledgments}
This work is supported by the National Natural Science Foundation of China (NNSFC) under Grant No.12475111. and No.12205387, and the Fundamental Research Funds for the Central Universities, Sun Yat-sen University.  

\section*{Note added}

While this work, motivated by the recent development in WI~\cite{Berghaus:2025dqi}, was nearing completion, ref.~\cite{Montefalcone:2025gxx} appeared on arXiv, which studied the thermal gravitons from warm inflation.

\end{acknowledgments}

\appendix
\section{UV freeze-in dark matter in warm inflation} 
\label{appA}
In this appendix, we briefly review the general aspects of UV freeze-in DM in WI. 

In the framework of UV freeze-in DM~\cite{UVDM}, the hidden and visible sectors are coupled only via non-renormalisable operators.The evolution of the number density of DM is governed by the following Boltzmann equation.
\begin{equation}\label{UV}
    \frac{d n_{\rm DM}}{dt}+3Hn_{\rm DM}=\mathcal{C}\frac{T^{2n+4}}{\Lambda^{2n}}
\end{equation}
The source term on the right-hand side represents the effective DM-radiation interaction arising from a non-renormalisable operator of mass dimension $n+4$, with $\Lambda$ being the cutoff scale of the effective description and $\mathcal{C}$ a dimensionless coefficient. For the example of purely gravitational DM discussed in section.~\ref{FIPPGDM}, $\Lambda$ is just the Planck mass and $n=2$, $\mathcal{C}=3096\zeta_3^2/5\pi^3$.

Using $dN\equiv Hdt$, the solution of eq.~\eqref{UV} reads:
\begin{equation}\label{DM solution}
    n_{\rm DM}(N)=e^{-3N}\int_{N_0}^NI_{\rm DM}(N')dN'
\end{equation}
where
\begin{equation}
    I_{\rm DM}(N)\equiv \mathcal{C}e^{3N}\frac{T^{2n+4}(N)}{\Lambda^{2n}H(N)}
\end{equation}
is the production rate of the comoving DM number density: $I_{\rm DM}=d(e^{3N}n_{\rm DM})/dN$. 

According to eq.~\eqref{DM solution}, the DM number density depends on the evolution of temperature and Hubble parameter, and therefore on the specific WI model. However, an overarching conclusion can still be derived as follows.

During the inflationary phase, both the inflaton and radiation undergo slow-roll evolution, causing the temperature $T(N)$ and the Hubble parameter $H(N)$ to remain approximately constant. Consequently, $I_{\rm DM}(N)$ becomes an exponentially increasing function. However, after inflation ends and the radiation-dominated era begins, it becomes exponentially decreasing, as $T(N)\propto e^{-N}$ and $H(N)\propto e^{-2N}$. Therefore, the function $I_{\rm DM}(N)$ is expected to exhibit a sharp peak at a specific e-fold, $N_{\rm peak}^{\rm DM}$, determined by $dI_{\rm DM}(N)/dN=0$. Explicitly, this condition reads:
\begin{equation}
    \left.\left[3+(2n+4)\frac{d\ln{T(N)}}{dN} - \frac{d\ln H(N)}{dN}\right]\right|_{N = N_{\mathrm{peak}}^{\rm DM}} = 0
\end{equation}
This property allows us to estimate the DM number density at some late times after $N_{\rm peak}^{\rm DM}$. We obtain:
\begin{equation}
n_{\rm DM}(N)\simeq e^{-3N}I_{\rm DM}(N_{\rm peak}^{\rm DM})\Delta N_{\mathrm{peak}}^{\rm DM}\quad(N>N_{\mathrm{peak}}^{\rm DM})
\end{equation}
with $\Delta N_{\mathrm{peak}}^{\rm DM}\gtrsim1$ being the half-width of $I_{\rm DM}(N)$.

The DM yield, $Y_{\rm DM}\equiv n_{\rm DM}/s$, can then be estimated~\cite{Freese:2024ogj},
\begin{equation}
Y_{\rm DM}(N)\simeq \frac{45}{2\pi^{2}g_{*s}}\frac{e^{3(N_{\mathrm{peak}}^{\rm DM}-N)}}{\Lambda^{2n}T^{3}(N)}\Delta N_{\mathrm{peak}}^{\rm DM}
\times\frac{ T^{2n+4}(N_{\mathrm{peak}}^{\rm DM})}{H(N_{\mathrm{peak}}^{\rm DM})}\quad(N>N_{\mathrm{peak}}^{\rm DM})
\end{equation}
To better understand this result, we compare it with the final DM yield produced during radiation dominated epoch with the reheating temperature $T_{RH}\equiv T(\epsilon_{H}=2)$:
\begin{equation}
    Y_{\rm DM}^{\mathrm{RH}} \simeq \frac{1}{4\sqrt{\pi}}\left(\frac{45}{\pi^2g_*}\right)^{\frac{3}{2}}\frac{1}{2n-1}\frac{M_{\rm pl}T_\mathrm{RH}^{2n-1}}{\Lambda^{2n}}
\end{equation}
In WI, DM production begins at the onset of inflation and always peaks before radiation domination. It follows that the DM yield is expected to be significantly larger than that which would have accumulated from the radiation-dominated epoch alone. This enhancement is then quantified by the ratio:
\begin{equation}
\frac{Y_{\rm DM}(N)}{Y_{\rm DM}^{\mathrm{RH}}}\simeq (2n-1)\frac{I_{\rm DM}(N_{\mathrm{peak}}^{\rm DM})}{I_{\rm DM}(N_{\mathrm{RH}})}\Delta N_{\mathrm{peak}}^{\rm DM}
\end{equation}

\begin{figure}[htbp]
	\centering
	\begin{minipage}{0.5\linewidth}
		\centering
		\includegraphics[width=1.\linewidth]{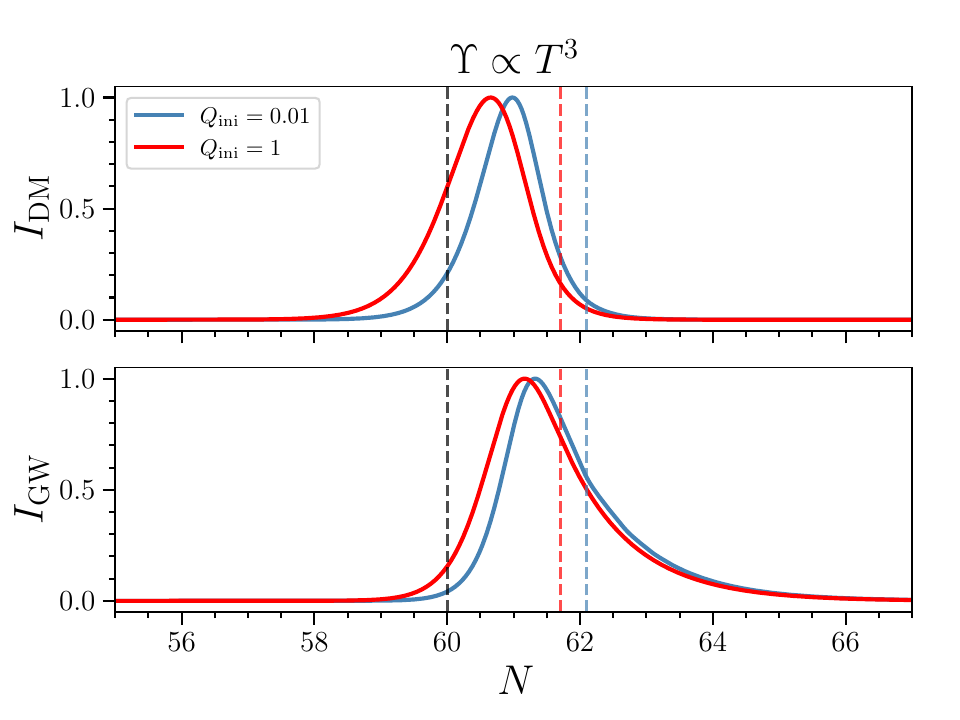}
	\end{minipage}%
	\begin{minipage}{0.5\linewidth}
		\centering
		\includegraphics[width=1.\linewidth]{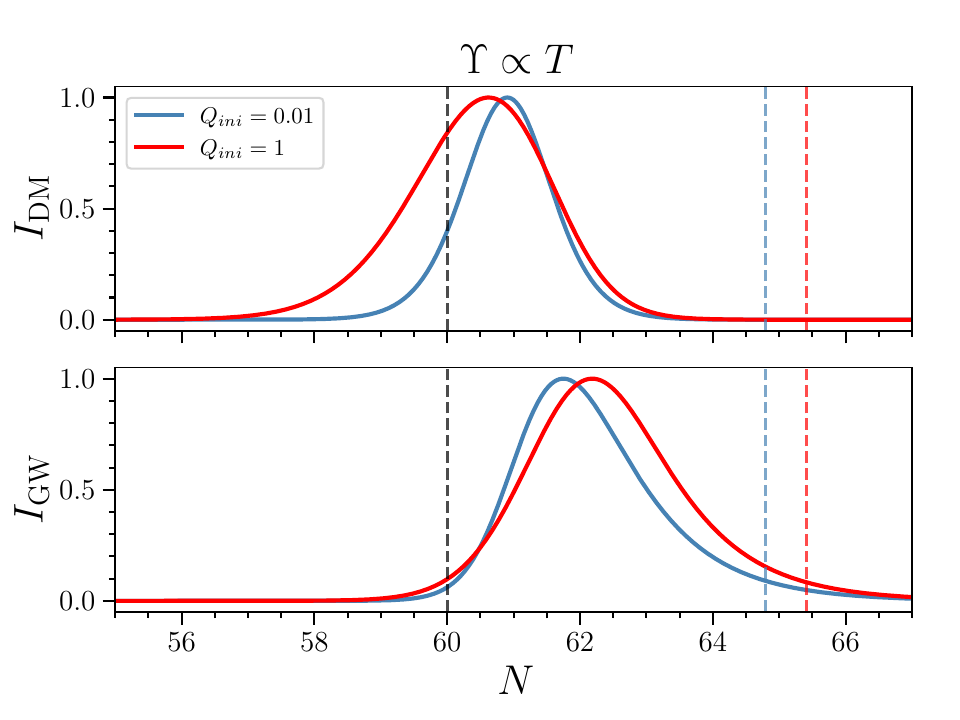}
	\end{minipage}
    \caption{The evolution of $I_{\mathrm{DM}}$ and $I_\mathrm{GW}$ in WI with dissipation term $\Upsilon\propto T^3$ and $\Upsilon\propto T$, respectively. All curves are normalized by their peak values. The blue line corresponds to $Q_{\mathrm{ini}}=0.01$ and the red line corresponds to $Q_{\mathrm{ini}}=1$. Vertical lines mark key transitions: the black dashed line indicates inflation termination ($\epsilon_{H}=1$), while the colored dashed lines show the start of radiation domination ($\epsilon_{H}=2$) for each $Q_{\rm ini}$ values. Note that $I_{\mathrm{DM}}$ and $I_\mathrm{GW}$ both sharply peaked at some e-folding numbers in the post-inflationary epoch, prior to the onset of radiation domination.}
    \label{fig:IDM and IGW}
\end{figure}

This comparison is equivalent to comparing the DM yield in WI with that in CI under the same reheating temperature $T_{\rm RH}$, since in the CI scenario, DM production occurs predominantly during the radiation-dominated era. And given $I_{\rm DM}(N_{\mathrm{peak}}^{\rm DM}) \gg I_{\rm DM}(N_{\rm RH})$ with $\Delta N_{\mathrm{peak}}^{\rm DM}\gtrsim1$,it follows that the relic abundance is significantly enhanced in WI compared to CI.

Figure~\ref{fig:IDM and IGW} shows the evolution of $I_\mathrm{DM}$ and $I_{\mathrm{GW}}$ in WI for both $\Upsilon \propto T^3$ and  $\Upsilon \propto T$, under both $Q_{\mathrm{ini}}=0.01$ and $Q_{\mathrm{ini}}=1$. We set $n=2$ since this is the case of purely gravitational DM as we have discussed in section.~\ref{FIPPGDM}. 
All curves are normalized to their respective peak values. The blue and red solid lines represent the cases with $Q_{\mathrm{ini}} = 0.01$ and $Q_{\mathrm{ini}} = 1$, respectively. The black dashed line marks the end of inflation, while the blue and red dashed lines indicate the onset of radiation domination for each scenario. 

As clearly shown in this figure and section.~\ref{FIPG}, graviton production bears notable similarities to the UV freeze-in of DM. Most graviton energy and DM particles are generated prior to the radiation-dominated epoch. The ratio of their respective production enhancements can be estimated on the order of $I_{X}(N^{X}_{\rm peak})/I_{X}(N_{RH})$, where $X$ stands for DM or GW.


\bibliographystyle{JHEP}
\bibliography{ref}

\providecommand{\href}[2]{#2}\begingroup\raggedright\begin{thebibliography}{10}

\bibitem{Berghaus:2025dqi}
K.V.~Berghaus, M.~Drewes and S.~Zell, \emph{{Warm Inflation with the Standard Model}}, \href{https://doi.org/10.1103/9nn9-bsm9}{\emph{Phys. Rev. Lett.} {\bfseries 135} (2025) 171002} [\href{https://arxiv.org/abs/2503.18829}{{\ttfamily 2503.18829}}].

\bibitem{STAROBINSKY198099}
A.~Starobinsky, \emph{A new type of isotropic cosmological models without singularity}, \href{https://doi.org/https://doi.org/10.1016/0370-2693(80)90670-X}{\emph{Physics Letters B} {\bfseries 91} (1980) 99}.

\bibitem{Guth1981}
A.H.~Guth, \emph{Inflationary universe: A possible solution to the horizon and flatness problems}, \href{https://doi.org/10.1103/PhysRevD.23.347}{\emph{Phys. Rev. D} {\bfseries 23} (1981) 347}.

\bibitem{LINDE1}
A.~Linde, \emph{A new inflationary universe scenario: A possible solution of the horizon, flatness, homogeneity, isotropy and primordial monopole problems}, \href{https://doi.org/https://doi.org/10.1016/0370-2693(82)91219-9}{\emph{Physics Letters B} {\bfseries 108} (1982) 389}.

\bibitem{WI1}
A.~Berera and L.-Z.~Fang, \emph{Thermally induced density perturbations in the inflation era}, \href{https://doi.org/10.1103/PhysRevLett.74.1912}{\emph{Phys. Rev. Lett.} {\bfseries 74} (1995) 1912}.

\bibitem{WI2}
A.~Berera, \emph{{Warm inflation}}, \href{https://doi.org/10.1103/PhysRevLett.75.3218}{\emph{Phys. Rev. Lett.} {\bfseries 75} (1995) 3218} [\href{https://arxiv.org/abs/astro-ph/9509049}{{\ttfamily astro-ph/9509049}}].

\bibitem{WI3}
A.~Berera, M.~Gleiser and R.O.~Ramos, \emph{A first principles warm inflation model that solves the cosmological horizon and flatness problems}, \href{https://doi.org/10.1103/PhysRevLett.83.264}{\emph{Phys. Rev. Lett.} {\bfseries 83} (1999) 264}.

\bibitem{WI4}
A.~Berera, I.G.~Moss and R.O.~Ramos, \emph{Warm inflation and its microphysical basis}, \href{https://doi.org/10.1088/0034-4885/72/2/026901}{\emph{Reports on Progress in Physics} {\bfseries 72} (2009) 026901}.

\bibitem{WI5}
V.~Kamali, M.~Motaharfar and R.~O.~Ramos, \emph{{Recent Developments in Warm Inflation}}, \href{https://doi.org/10.3390/universe9030124}{\emph{Universe} {\bfseries 9} (2023) 124} [\href{https://arxiv.org/abs/2302.02827}{{\ttfamily 2302.02827}}].

\bibitem{huang2025particles}
F.P.~Huang, \emph{{The First Particles}},  1, 2025.

\bibitem{Yokoyama:1998ju}
J.~Yokoyama and A.D.~Linde, \emph{{Is warm inflation possible?}}, \href{https://doi.org/10.1103/PhysRevD.60.083509}{\emph{Phys. Rev. D} {\bfseries 60} (1999) 083509} [\href{https://arxiv.org/abs/hep-ph/9809409}{{\ttfamily hep-ph/9809409}}].

\bibitem{SMWI2}
R.~O.~Ramos and G.S.~Rodrigues, \emph{{Viability of warm inflation with standard model interactions}}, \href{https://doi.org/10.1103/wn1m-19gt}{\emph{Phys. Rev. D} {\bfseries 111} (2025) 123527} [\href{https://arxiv.org/abs/2504.20943}{{\ttfamily 2504.20943}}].

\bibitem{Preheat1997}
S.Y.~Khlebnikov and I.I.~Tkachev, \emph{{Relic gravitational waves produced after preheating}}, \href{https://doi.org/10.1103/PhysRevD.56.653}{\emph{Phys. Rev. D} {\bfseries 56} (1997) 653} [\href{https://arxiv.org/abs/hep-ph/9701423}{{\ttfamily hep-ph/9701423}}].

\bibitem{PreheatGW2}
B.~Bassett, \emph{{The Preheating - gravitational wave correspondence}}, \href{https://doi.org/10.1103/PhysRevD.56.3439}{\emph{Phys. Rev. D} {\bfseries 56} (1997) 3439} [\href{https://arxiv.org/abs/hep-ph/9704399}{{\ttfamily hep-ph/9704399}}].

\bibitem{PreheatGW2007}
J.~Garcia-Bellido and D.G.~Figueroa, \emph{{A stochastic background of gravitational waves from hybrid preheating}}, \href{https://doi.org/10.1103/PhysRevLett.98.061302}{\emph{Phys. Rev. Lett.} {\bfseries 98} (2007) 061302} [\href{https://arxiv.org/abs/astro-ph/0701014}{{\ttfamily astro-ph/0701014}}].

\bibitem{PreheatGW2008}
J.~Garcia-Bellido, D.G.~Figueroa and A.~Sastre, \emph{{A Gravitational Wave Background from Reheating after Hybrid Inflation}}, \href{https://doi.org/10.1103/PhysRevD.77.043517}{\emph{Phys. Rev. D} {\bfseries 77} (2008) 043517} [\href{https://arxiv.org/abs/0707.0839}{{\ttfamily 0707.0839}}].

\bibitem{ReheatGW2019}
K.~Nakayama and Y.~Tang, \emph{{Stochastic Gravitational Waves from Particle Origin}}, \href{https://doi.org/10.1016/j.physletb.2018.11.023}{\emph{Phys. Lett. B} {\bfseries 788} (2019) 341} [\href{https://arxiv.org/abs/1810.04975}{{\ttfamily 1810.04975}}].

\bibitem{ReheatGW2}
D.~Huang and L.~Yin, \emph{{Stochastic Gravitational Waves from Inflaton Decays}}, \href{https://doi.org/10.1103/PhysRevD.100.043538}{\emph{Phys. Rev. D} {\bfseries 100} (2019) 043538} [\href{https://arxiv.org/abs/1905.08510}{{\ttfamily 1905.08510}}].

\bibitem{Barman:2023ymn}
B.~Barman, N.~Bernal, Y.~~ and {\'O}.~Zapata, \emph{{Gravitational wave from graviton Bremsstrahlung during reheating}}, \href{https://doi.org/10.1088/1475-7516/2023/05/019}{\emph{JCAP} {\bfseries 05} (2023) 019} [\href{https://arxiv.org/abs/2301.11345}{{\ttfamily 2301.11345}}].

\bibitem{Datta:2024tne}
A.~Datta and A.~Sil, \emph{{Probing Leptogenesis through Gravitational Waves}},  10, 2024.

\bibitem{Datta:2025wfh}
A.~Datta, S.~Khalil, R.K.~Mandal and A.~Sil, \emph{{Probing Right Handed Neutrino assisted Reheating with Gravitational Waves and Leptogenesis}},  7, 2025.

\bibitem{Xu:2025wjq}
X.-J.~Xu, Y.~Xu, Q.~Yin and J.~Zhu, \emph{{Full-spectrum analysis of gravitational wave production from inflation to reheating}}, \href{https://doi.org/10.1007/JHEP10(2025)141}{\emph{JHEP} {\bfseries 10} (2025) 141} [\href{https://arxiv.org/abs/2505.08868}{{\ttfamily 2505.08868}}].

\bibitem{defect1}
A.~Vilenkin, \emph{Gravitational radiation from cosmic strings}, \href{https://doi.org/https://doi.org/10.1016/0370-2693(81)91144-8}{\emph{Physics Letters B} {\bfseries 107} (1981) 47}.

\bibitem{defect2}
L.M.~Krauss, \emph{Gravitational waves from global phase transitions}, \href{https://doi.org/https://doi.org/10.1016/0370-2693(92)90425-4}{\emph{Physics Letters B} {\bfseries 284} (1992) 229}.

\bibitem{defect3}
C.J.~Hogan and M.J.~Rees, \emph{{Gravitational interactions of cosmic strings}}, \href{https://doi.org/10.1038/311109a0}{\emph{Nature} {\bfseries 311} (1984) 109}.

\bibitem{phase1}
C.J.~Hogan, \emph{{Gravitational radiation from cosmological phase transitions}}, \href{https://doi.org/10.1093/mnras/218.4.629}{\emph{Mon. Not. Roy. Astron. Soc.} {\bfseries 218} (1986) 629}.

\bibitem{phase2}
A.~Kosowsky, M.S.~Turner and R.~Watkins, \emph{{Gravitational radiation from colliding vacuum bubbles}}, \href{https://doi.org/10.1103/PhysRevD.45.4514}{\emph{Phys. Rev. D} {\bfseries 45} (1992) 4514}.

\bibitem{phase3}
M.~Kamionkowski, A.~Kosowsky and M.S.~Turner, \emph{{Gravitational radiation from first order phase transitions}}, \href{https://doi.org/10.1103/PhysRevD.49.2837}{\emph{Phys. Rev. D} {\bfseries 49} (1994) 2837} [\href{https://arxiv.org/abs/astro-ph/9310044}{{\ttfamily astro-ph/9310044}}].

\bibitem{phase4}
C.~Caprini et~al., \emph{{Science with the space-based interferometer eLISA. II: Gravitational waves from cosmological phase transitions}}, \href{https://doi.org/10.1088/1475-7516/2016/04/001}{\emph{JCAP} {\bfseries 04} (2016) 001} [\href{https://arxiv.org/abs/1512.06239}{{\ttfamily 1512.06239}}].

\bibitem{phase5}
T.~Konstandin, \emph{{Gravitational radiation from a bulk flow model}}, \href{https://doi.org/10.1088/1475-7516/2018/03/047}{\emph{JCAP} {\bfseries 03} (2018) 047} [\href{https://arxiv.org/abs/1712.06869}{{\ttfamily 1712.06869}}].

\bibitem{phase6}
M.~Hindmarsh and M.~Hijazi, \emph{{Gravitational waves from first order cosmological phase transitions in the Sound Shell Model}}, \href{https://doi.org/10.1088/1475-7516/2019/12/062}{\emph{JCAP} {\bfseries 12} (2019) 062} [\href{https://arxiv.org/abs/1909.10040}{{\ttfamily 1909.10040}}].

\bibitem{turbulence1}
A.~Kosowsky, A.~Mack and T.~Kahniashvili, \emph{Gravitational radiation from cosmological turbulence}, \href{https://doi.org/10.1103/PhysRevD.66.024030}{\emph{Phys. Rev. D} {\bfseries 66} (2002) 024030}.

\bibitem{turbulence2}
A.~Nicolis, \emph{{Relic gravitational waves from colliding bubbles and cosmic turbulence}}, \href{https://doi.org/10.1088/0264-9381/21/4/L05}{\emph{Class. Quant. Grav.} {\bfseries 21} (2004) L27} [\href{https://arxiv.org/abs/gr-qc/0303084}{{\ttfamily gr-qc/0303084}}].

\bibitem{turbulence3}
S.J.~Huber and T.~Konstandin, \emph{{Gravitational Wave Production by Collisions: More Bubbles}}, \href{https://doi.org/10.1088/1475-7516/2008/09/022}{\emph{JCAP} {\bfseries 09} (2008) 022} [\href{https://arxiv.org/abs/0806.1828}{{\ttfamily 0806.1828}}].

\bibitem{turbulence4}
C.~Caprini, R.~Durrer and G.~Servant, \emph{{The stochastic gravitational wave background from turbulence and magnetic fields generated by a first-order phase transition}}, \href{https://doi.org/10.1088/1475-7516/2009/12/024}{\emph{JCAP} {\bfseries 12} (2009) 024} [\href{https://arxiv.org/abs/0909.0622}{{\ttfamily 0909.0622}}].

\bibitem{Ghiglieri:2015nfa}
J.~Ghiglieri and M.~Laine, \emph{{Gravitational wave background from Standard Model physics: Qualitative features}}, \href{https://doi.org/10.1088/1475-7516/2015/07/022}{\emph{JCAP} {\bfseries 07} (2015) 022} [\href{https://arxiv.org/abs/1504.02569}{{\ttfamily 1504.02569}}].

\bibitem{Ghiglieri:2020mhm}
J.~Ghiglieri, G.~Jackson, M.~Laine and Y.~Zhu, \emph{{Gravitational wave background from Standard Model physics: Complete leading order}}, \href{https://doi.org/10.1007/JHEP07(2020)092}{\emph{JHEP} {\bfseries 07} (2020) 092} [\href{https://arxiv.org/abs/2004.11392}{{\ttfamily 2004.11392}}].

\bibitem{Ringwald:2020ist}
A.~Ringwald, J.~Sch{\"u}tte-Engel and C.~Tamarit, \emph{{Gravitational Waves as a Big Bang Thermometer}}, \href{https://doi.org/10.1088/1475-7516/2021/03/054}{\emph{JCAP} {\bfseries 03} (2021) 054} [\href{https://arxiv.org/abs/2011.04731}{{\ttfamily 2011.04731}}].

\bibitem{Ghiglieri:2022rfp}
J.~Ghiglieri, J.~Sch{\"u}tte-Engel and E.~Speranza, \emph{{Freezing-in gravitational waves}}, \href{https://doi.org/10.1103/PhysRevD.109.023538}{\emph{Phys. Rev. D} {\bfseries 109} (2024) 023538} [\href{https://arxiv.org/abs/2211.16513}{{\ttfamily 2211.16513}}].

\bibitem{Drewes:2023oxg}
M.~Drewes, Y.~Georis, J.~Klaric and P.~Klose, \emph{{Upper bound on thermal gravitational wave backgrounds from hidden sectors}}, \href{https://doi.org/10.1088/1475-7516/2024/06/073}{\emph{JCAP} {\bfseries 06} (2024) 073} [\href{https://arxiv.org/abs/2312.13855}{{\ttfamily 2312.13855}}].

\bibitem{2graviton}
J.~Ghiglieri, M.~Laine, J.~Sch{\"u}tte-Engel and E.~Speranza, \emph{{Double-graviton production from Standard Model plasma}}, \href{https://doi.org/10.1088/1475-7516/2024/04/062}{\emph{JCAP} {\bfseries 04} (2024) 062} [\href{https://arxiv.org/abs/2401.08766}{{\ttfamily 2401.08766}}].

\bibitem{Montefalcone:2025gxx}
G.~Montefalcone, B.~Shams Es~Haghi, T.~~ and K.~Freese, \emph{{Thermal gravitons from warm inflation}}, \href{https://doi.org/10.1103/rnvb-t4lx}{\emph{Phys. Rev. D} {\bfseries 112} (2025) 063556} [\href{https://arxiv.org/abs/2507.08739}{{\ttfamily 2507.08739}}].

\bibitem{UVDM}
F.~Elahi, C.~Kolda and J.~Unwin, \emph{{UltraViolet Freeze-in}}, \href{https://doi.org/10.1007/JHEP03(2015)048}{\emph{JHEP} {\bfseries 03} (2015) 048} [\href{https://arxiv.org/abs/1410.6157}{{\ttfamily 1410.6157}}].

\bibitem{Freese:2024ogj}
K.~Freese, G.~Montefalcone and B.~Shams Es~Haghi, \emph{{Dark Matter Production during Warm Inflation via Freeze-In}}, \href{https://doi.org/10.1103/PhysRevLett.133.211001}{\emph{Phys. Rev. Lett.} {\bfseries 133} (2024) 211001} [\href{https://arxiv.org/abs/2401.17371}{{\ttfamily 2401.17371}}].

\bibitem{WIFIDM}
R.~de~Souza, J.G.~Rodrigues, C.~Siqueira, F.B.M.d.~Santos and J.~Alcaniz, \emph{{Dark Matter freeze-in during warm inflation and the seesaw mechanism}}, \href{https://doi.org/10.1007/JHEP04(2025)125}{\emph{JHEP} {\bfseries 04} (2025) 125} [\href{https://arxiv.org/abs/2412.06778}{{\ttfamily 2412.06778}}].

\bibitem{Berghaus:2019whh}
K.V.~Berghaus, P.W.~Graham and D.E.~Kaplan, \emph{{Minimal Warm Inflation}}, \href{https://doi.org/10.1088/1475-7516/2020/03/034}{\emph{JCAP} {\bfseries 03} (2020) 034} [\href{https://arxiv.org/abs/1910.07525}{{\ttfamily 1910.07525}}].

\bibitem{Laine:2021ego}
M.~Laine and S.~Procacci, \emph{{Minimal warm inflation with complete medium response}}, \href{https://doi.org/10.1088/1475-7516/2021/06/031}{\emph{JCAP} {\bfseries 06} (2021) 031} [\href{https://arxiv.org/abs/2102.09913}{{\ttfamily 2102.09913}}].

\bibitem{sph1}
G.D.~Moore and M.~Tassler, \emph{{The Sphaleron Rate in SU(N) Gauge Theory}}, \href{https://doi.org/10.1007/JHEP02(2011)105}{\emph{JHEP} {\bfseries 02} (2011) 105} [\href{https://arxiv.org/abs/1011.1167}{{\ttfamily 1011.1167}}].

\bibitem{Bastero-Gil:2016qru}
M.~Bastero-Gil, A.~Berera, R.O.~Ramos and J.G.~Rosa, \emph{{Warm Little Inflaton}}, \href{https://doi.org/10.1103/PhysRevLett.117.151301}{\emph{Phys. Rev. Lett.} {\bfseries 117} (2016) 151301} [\href{https://arxiv.org/abs/1604.08838}{{\ttfamily 1604.08838}}].

\bibitem{WI2easy}
G.S.~Rodrigues and R.~O.~Ramos, \emph{{WI2easy: warm inflation dynamics made easy}}, \href{https://doi.org/10.1088/1475-7516/2025/09/014}{\emph{JCAP} {\bfseries 09} (2025) 014} [\href{https://arxiv.org/abs/2504.17760}{{\ttfamily 2504.17760}}].

\bibitem{Planck:2018jri}
{\scshape Planck} collaboration, \emph{{Planck 2018 results. X. Constraints on inflation}}, \href{https://doi.org/10.1051/0004-6361/201833887}{\emph{Astron. Astrophys.} {\bfseries 641} (2020) A10} [\href{https://arxiv.org/abs/1807.06211}{{\ttfamily 1807.06211}}].

\bibitem{Planck:2018vyg}
{\scshape Planck} collaboration, \emph{{Planck 2018 results. VI. Cosmological parameters}}, \href{https://doi.org/10.1051/0004-6361/201833910}{\emph{Astron. Astrophys.} {\bfseries 641} (2020) A6} [\href{https://arxiv.org/abs/1807.06209}{{\ttfamily 1807.06209}}].

\bibitem{LISA:2017pwj}
{\scshape LISA} collaboration, P.~Amaro-Seoane et~al., \emph{{Laser Interferometer Space Antenna}},  2, 2017.

\bibitem{TianQin:2015yph}
{\scshape TianQin} collaboration, \emph{{TianQin: a space-borne gravitational wave detector}}, \href{https://doi.org/10.1088/0264-9381/33/3/035010}{\emph{Class. Quant. Grav.} {\bfseries 33} (2016) 035010} [\href{https://arxiv.org/abs/1512.02076}{{\ttfamily 1512.02076}}].

\bibitem{Hu:2017mde}
W.-R.~Hu and Y.-L.~Wu, \emph{{The Taiji Program in Space for gravitational wave physics and the nature of gravity}}, \href{https://doi.org/10.1093/nsr/nwx116}{\emph{Natl. Sci. Rev.} {\bfseries 4} (2017) 685}.

\bibitem{Crowder:2005nr}
J.~Crowder and N.J.~Cornish, \emph{{Beyond LISA: Exploring future gravitational wave missions}}, \href{https://doi.org/10.1103/PhysRevD.72.083005}{\emph{Phys. Rev. D} {\bfseries 72} (2005) 083005} [\href{https://arxiv.org/abs/gr-qc/0506015}{{\ttfamily gr-qc/0506015}}].

\bibitem{Corbin:2005ny}
V.~Corbin and N.J.~Cornish, \emph{{Detecting the cosmic gravitational wave background with the big bang observer}}, \href{https://doi.org/10.1088/0264-9381/23/7/014}{\emph{Class. Quant. Grav.} {\bfseries 23} (2006) 2435} [\href{https://arxiv.org/abs/gr-qc/0512039}{{\ttfamily gr-qc/0512039}}].

\bibitem{Reitze:2019iox}
D.~Reitze et~al., \emph{{Cosmic Explorer: The U.S. Contribution to Gravitational-Wave Astronomy beyond LIGO}}, {\emph{Bull. Am. Astron. Soc.} {\bfseries 51} (2019) 035} [\href{https://arxiv.org/abs/1907.04833}{{\ttfamily 1907.04833}}].

\bibitem{Sathyaprakash:2012jk}
B.~Sathyaprakash et~al., \emph{{Scientific Objectives of Einstein Telescope}}, \href{https://doi.org/10.1088/0264-9381/29/12/124013}{\emph{Class. Quant. Grav.} {\bfseries 29} (2012) 124013} [\href{https://arxiv.org/abs/1206.0331}{{\ttfamily 1206.0331}}].

\bibitem{ET:2019dnz}
{\scshape ET} collaboration, \emph{{Science Case for the Einstein Telescope}}, \href{https://doi.org/10.1088/1475-7516/2020/03/050}{\emph{JCAP} {\bfseries 03} (2020) 050} [\href{https://arxiv.org/abs/1912.02622}{{\ttfamily 1912.02622}}].

\bibitem{Herman:2022fau}
N.~Herman, L.~Lehoucq and A.~F{\'{u}}zfa, \emph{{Electromagnetic antennas for the resonant detection of the stochastic gravitational wave background}}, \href{https://doi.org/10.1103/PhysRevD.108.124009}{\emph{Phys. Rev. D} {\bfseries 108} (2023) 124009} [\href{https://arxiv.org/abs/2203.15668}{{\ttfamily 2203.15668}}].

\bibitem{GDMTang:2016}
Y.~Tang and Y.-L.~Wu, \emph{{Pure Gravitational Dark Matter, Its Mass and Signatures}}, \href{https://doi.org/10.1016/j.physletb.2016.05.045}{\emph{Phys. Lett. B} {\bfseries 758} (2016) 402} [\href{https://arxiv.org/abs/1604.04701}{{\ttfamily 1604.04701}}].

\bibitem{Garny:2017kha}
M.~Garny, A.~Palessandro, M.~Sandora and M.S.~Sloth, \emph{{Theory and Phenomenology of Planckian Interacting Massive Particles as Dark Matter}}, \href{https://doi.org/10.1088/1475-7516/2018/02/027}{\emph{JCAP} {\bfseries 02} (2018) 027} [\href{https://arxiv.org/abs/1709.09688}{{\ttfamily 1709.09688}}].

\bibitem{GDMEma:2018ucl}
Y.~Ema, K.~Nakayama and Y.~Tang, \emph{{Production of Purely Gravitational Dark Matter}}, \href{https://doi.org/10.1007/JHEP09(2018)135}{\emph{JHEP} {\bfseries 09} (2018) 135} [\href{https://arxiv.org/abs/1804.07471}{{\ttfamily 1804.07471}}].

\bibitem{Tang:2017hvq}
Y.~Tang and Y.-L.~Wu, \emph{{On Thermal Gravitational Contribution to Particle Production and Dark Matter}}, \href{https://doi.org/10.1016/j.physletb.2017.10.034}{\emph{Phys. Lett. B} {\bfseries 774} (2017) 676} [\href{https://arxiv.org/abs/1708.05138}{{\ttfamily 1708.05138}}].

\bibitem{Bernal:2018qlk}
N.~Bernal, M.~Dutra, Y.~Mambrini, K.~Olive, M.~Peloso and M.~Pierre, \emph{{Spin-2 Portal Dark Matter}}, \href{https://doi.org/10.1103/PhysRevD.97.115020}{\emph{Phys. Rev. D} {\bfseries 97} (2018) 115020} [\href{https://arxiv.org/abs/1803.01866}{{\ttfamily 1803.01866}}].

\bibitem{GDMinCIHashiba:2018tbu}
S.~Hashiba and J.~Yokoyama, \emph{{Gravitational particle creation for dark matter and reheating}}, \href{https://doi.org/10.1103/PhysRevD.99.043008}{\emph{Phys. Rev. D} {\bfseries 99} (2019) 043008} [\href{https://arxiv.org/abs/1812.10032}{{\ttfamily 1812.10032}}].

\bibitem{GDMinCIMambrini:2021zpp}
Y.~Mambrini and K.A.~Olive, \emph{{Gravitational Production of Dark Matter during Reheating}}, \href{https://doi.org/10.1103/PhysRevD.103.115009}{\emph{Phys. Rev. D} {\bfseries 103} (2021) 115009} [\href{https://arxiv.org/abs/2102.06214}{{\ttfamily 2102.06214}}].

\bibitem{Clery:2021bwz}
S.~Clery, Y.~Mambrini, K.A.~Olive and S.~Verner, \emph{{Gravitational portals in the early Universe}}, \href{https://doi.org/10.1103/PhysRevD.105.075005}{\emph{Phys. Rev. D} {\bfseries 105} (2022) 075005} [\href{https://arxiv.org/abs/2112.15214}{{\ttfamily 2112.15214}}].

\bibitem{wqy2025}
Q.-Y.~Wang, T.~Jia, P.-R.~Chen and Y.~Tang, \emph{{Purely gravitational dark matter production in warm inflation}}, \href{https://doi.org/10.1103/jdkz-8z54}{\emph{Phys. Rev. D} {\bfseries 112} (2025) 043508} [\href{https://arxiv.org/abs/2504.13147}{{\ttfamily 2504.13147}}].

\end{thebibliography}\endgroup

\end{document}